\documentclass[11pt,preprint]{aastex61}

\usepackage{natbib}
\usepackage {float}
\usepackage{graphicx}
\usepackage{amssymb,amsmath,mathtools}

\received{}
\accepted{}
\shorttitle{H$_0$ from {\it Gaia} EDR3 and MW Cepheids}
\shortauthors{Riess et al.}

\newcommand{\xddots}{%
  \raise 5pt \hbox {.}
  \mkern 6mu
  \raise 1pt \hbox {.}
  \mkern 6mu
  \raise -3pt \hbox {.}
}

\newcommand{\kmsmpc}{\hbox{$ \, \rm km\, s^{-1} \, Mpc^{-1}$}}

\newcommand{\bq}{\begin{equation}} 
\newcommand{\eq}{\end{equation}}   
\newcommand{\hubbleconst}{\hbox{H$_0$}}

\newcommand{\hoedr}{\hbox{$ 73.0 \pm 1.4$}}  % \, \rm km\, s^{-1} \, Mpc^{-1}$}} 
\newcommand{\hoall}{\hbox{$ 73.2 \pm 1.3$}} %  \, \rm km\, s^{-1} \, Mpc^{-1}$}} 
\newcommand{\hoedrss}{\hbox{$ 73.2 \pm 1.4 $}} % \, \rm km\, s^{-1} \, Mpc^{-1}$}} 

\newcommand{\mias}{$ \,\mu$as \,}  % microarcseconds
\newcommand{\muas}{\hbox{$\, \mu\rm as$}}
\newcommand{\mas}{\hbox{$\, \rm mas$}}

\newcommand{\fourzp}{$ -14 \pm 6 $ } % \,\mu$as}
\newcommand{\fourslope}{$-3.28 \pm 0.06$} 
\newcommand{\fourmetal}{$-0.20 \pm 0.13$ } % mag dex$^{-1}$} 
\newcommand{\fourchi}{68.0}

\newcommand{\fourzpwss}{$ -15 \pm 6 $} % \,\mu$as}
\newcommand{\fourslopewss}{$-3.34 \pm 0.06$} 
\newcommand{\fourmetalwss}{$-0.18 \pm 0.13$ } %  mag dex$^{-1}$} 
\newcommand{\fourchiwss}{78.8}

\newcommand{\twozp}{$ -14 \pm 6 $} % \,\mu$as}
\newcommand{\twomag}{$-5.915 \pm 0.022$}
\newcommand{\twochi}{68.2}

\newcommand{\onemag}{$-5.865 \pm 0.013$}
\newcommand{\onechi}{74.5}
\newcommand{\hoedrone}{\hbox{$ 74.7 \pm 1.3$}} %  \, \rm km\, s^{-1} \, Mpc^{-1}$}} 

\newcommand{\twomagzinn}{$-5.925 \pm 0.018$}

\newcommand{\twozpwss}{$ -15 \pm 6 $} % \,\mu$as}
\newcommand{\twomagwss}{$-5.910 \pm 0.022$}
\newcommand{\twochiwss}{81.2}

\newcommand{\faceplanck}{4.2} %73.2 +/- 1.3 vs 67.5 +/- 0.4

\newcommand{\beq}{\begin{equation}}
\newcommand{\eeq}{\end{equation}}
\newcommand{\beqa}{\begin{eqnarray}}
\newcommand{\eeqa}{\end{eqnarray}}

\newcommand{\PL}{$P\hbox{--}L$\ }
\newcommand{\PLs}{$P\hbox{--}L$}
\newcommand{\nd}{\multicolumn{1}{c}{$\dots$}}

\newcommand{\HST}{{\it HST}}
\newcommand{\Gaia}{{\it Gaia}}

\long\def\check#1{}
\long\def\hide#1{}

\begin{document} 

\title{Cosmic Distances Calibrated to 1\% Precision with {\it Gaia} EDR3 Parallaxes and {\it Hubble} Space Telescope Photometry of 75 Milky Way Cepheids Confirm Tension with $\Lambda$CDM}

\author{Adam G.~Riess}
\affiliation{Space Telescope Science Institute, 3700 San Martin Drive, Baltimore, MD 21218, USA}
\affiliation{Department of Physics and Astronomy, Johns Hopkins University, Baltimore, MD 21218, USA}

\author{Stefano Casertano}
\affiliation{Space Telescope Science Institute, 3700 San Martin Drive, Baltimore, MD 21218, USA}

\author{Wenlong Yuan}
\affiliation{Department of Physics and Astronomy, Johns Hopkins University, Baltimore, MD 21218, USA}

\author{J.~Bradley Bowers}
\affiliation{Department of Physics and Astronomy, Johns Hopkins University, Baltimore, MD 21218, USA}

\author{Lucas Macri}
\affiliation{Mitchell Institute for Fundamental Physics \& Astronomy, Department of Physics \& Astronomy,\\Texas A\&M University, College Station, TX 77843, USA}

\author{Joel C. Zinn}
\altaffiliation{NSF Astronomy and Astrophysics Postdoctoral Fellow.}
\affiliation{Department of Astrophysics, American Museum of Natural History, Central Park West at 79th Street,\\New York, NY 10024, USA}

\author{Dan Scolnic}
\affiliation{Department of Physics, Duke University, Durham, NC 27708, USA}

\begin{abstract} 
We present an expanded sample of 75 Milky Way Cepheids with {\it Hubble Space Telescope (HST)} photometry and {\Gaia} EDR3 parallaxes which we use to recalibrate the extragalactic distance ladder and refine the determination of the Hubble constant. All {\it HST} observations were obtained with the same instrument (WFC3) and filters ({\it F555W}, {\it F814W}, {\it F160W}) used for imaging of extragalactic Cepheids in Type Ia supernova (SN Ia) hosts. The {\it HST} observations used the WFC3 spatial scanning mode to mitigate saturation and reduce pixel-to-pixel calibration errors, reaching a mean photometric error of 5 millimags per observation. We use new {\Gaia} EDR3 parallaxes, vastly improved since DR2, and the Period-Luminosity ({\PLs}) relation of these Cepheids to simultaneously calibrate the extragalactic distance ladder and to refine the determination of the {\Gaia} EDR3 parallax offset. The resulting geometric calibration of Cepheid luminosities has 1.0\% precision, better than any alternative geometric anchor. Applied to the calibration of SNe~Ia, it results in a measurement of the Hubble constant of \hoedr \kmsmpc, in good agreement with conclusions based on earlier {\Gaia} data releases. We also find the slope of the Cepheid \PL relation in the Milky Way, and the metallicity dependence of its zeropoint, to be in good agreement with the mean values derived from other galaxies. In combination with the best complementary sources of Cepheid calibration, we reach 1.8\% precision and find $H_0=$\hoall \kmsmpc, a \faceplanck $\sigma$ difference with the prediction from {\it Planck} CMB observations under $\Lambda$CDM. We expect to reach $\sim$ 1.3\% precision in the near term from an expanded sample of $\sim$ 40 SNe Ia in Cepheid hosts.

\end{abstract} 

\keywords{astrometry: parallaxes --- cosmology: distance scale --- cosmology:
observations --- stars: variables: Cepheids --- supernovae: general}

\section{Introduction} 

This is the second paper in a series reporting on efforts to improve knowledge of the distance scale and the Hubble constant (\hubbleconst) by combining parallax measurements of Milky Way (MW) Cepheids from the ESA {\Gaia} mission \citep{Gaia-Collaboration:2016,Gaia-Collaboration:2016a, Gaia-Collaboration:2018} and multi-band photometry of these variables from the {\it Hubble} Space Telescope ({\HST}). This combination offers the only means at present to provide a $\sim\!1\% $ calibration of the Extragalactic Distance Scale. Neither facility can achieve this ambitious goal alone.  Reaching this milestone requires simultaneously measuring Cepheid mean parallaxes to $\sim\!5 \, \mu$as precision from {\Gaia} and measuring the mean brightness of the same objects to $\sim\!0.01$ mag precision with {\HST} on the {\it same} photometric systems used to measure their extragalactic counterparts. By using such purely {\it differential} flux measurements of Cepheids along the distance ladder, it is possible to circumvent systematic uncertainties related to zeropoints and transmission functions which otherwise incur a {\it systematic} uncertainty of $\sim 2-3$\% in the determination of H$_0$, nearly twice the target goal, even before considering any additional statistical and systematic uncertainties along the distance ladder. 

We started building this photometric bridge in 2012, by observing 50 MW Cepheid ``Standards'', randomly chosen by the {\HST} scheduling process among the 70 known Cepheids with periods of $P\!>\!8$~days, $H$-band extinction of $A_H\!<\!0.4 $~mag and $V\!>\!6$~mag that were targeted in our Cycle 20 SNAP program. These selection criteria were adopted to collect the most useful sample for calibration of thousands of extragalactic Cepheids observed in the hosts of 19 Type Ia supernovae (SNe~Ia) and the megamaser host NGC$\,$4258.  For all 50 Cepheids, we have published \citep[][hereafter R18a,b]{Riess:2018a,Riess:2018b} near-infrared (NIR) photometry collected with {\HST}/WFC3-IR in {\it F160W} (similar to the $H$ band) to reduce systematics caused by reddening and metallicity, and optical photometry obtained with {\HST}/WFC3-UVIS and/or ACS in {\it F555W} and {\it F814W} (similar to the $V$ and $I$ bands). Observations in these three filters can be combined to form a reddening-free distance measure \citep[][hereafter R16]{Hoffmann:2016,Riess:2016}.

Observing MW Cepheids, which are 15 to 20 astronomical magnitudes brighter than their extragalactic counterparts, requires extremely brief and precisely-known exposure times to mitigate saturation and provide accurate photometry. We accomplished this (R18a,b) through very fast spatial scans with {\HST}, moving the telescope during the observation so that the target covers a long, nearly vertical line over the detector. We used a scan speed of $ 7\farcs 5 / {\rm sec} $, corresponding to an effective exposure time of 5~ms in the visible and 20~ms in the infrared, much shorter than the minimum effective exposure times possible with the WFC3 hardware. Scanning observations are also free from the variations and uncertainties in shutter flight time (for {\it F555W} and {\it F814W} with WFC3-UVIS) that affect very short pointed observations \citep{Sahu:2015}. Spatial scans offer the additional advantage of varying the position of the source on the detector, which averages down pixel-to-pixel errors in the flat fields, and can also be used to vary the pixel phase, reducing the uncertainty from undersampled point-spread-function photometry. Finally, unlike ground-based photometry which relies on calibrators in the same region of the sky, {\HST} can measure the photometry of MW Cepheids over the whole sky, without concern about regional variations in calibrators. The original sample of 50 should have been sufficient to produce a $\sim$ 1\% precise calibration of the Cepheid Period-Luminosity relation (\PLs, often referred to as the Leavitt Law) in the absence of unexpected systematics in the {\Gaia} parallaxes.

Unfortunately, after launch {\Gaia} was found to suffer from a large thermal oscillation which produces a variation in the angle between its two fields of view and limits its ability to determine {\it absolute} parallaxes from relative astrometry measured at multiple parallax phases. The result is that {\Gaia} parallaxes are affected by an additive error, the so-called ``parallax zero point'', clearly identified in Data Release 2 \citep[DR2]{Gaia-Collaboration:2018} using quasars \citep[][hereafter L18]{Lindegren:2018}.  L18 attempted to characterize this error, but---possibly because the solution relies on a still imperfect focal plane calibration (see Fig.~16 in L18)---the additive term has been found to vary with the color and/or magnitude of the source and its location on the sky; for this reason, we refer to this term as ``parallax offset'', rather than zero point, as it is not a single value. Recognizing this, the {\Gaia} team recommended that users calibrate the offset applicable for their targets from sources with similar color and magnitude. Regrettably, quasars are much bluer and 5-10 mag fainter than MW Cepheids, making them unsuitable as offset calibrators. \citet{Riess:2018b} showed that the parallax offset could be measured directly from the Cepheids, but at a high cost in precision, increasing the uncertainty in H$_0$ by a factor of 2.5---from 1.3\% to 3.3\%. This lesser precision was sufficient to confirm the present ``H$_0$ tension,'' but is woefully short of what is possible with fully-offset-calibrated {\Gaia} parallaxes. However, with the {\Gaia} offset issue now fully recognized, {\HST} observations of additional MW Cepheids could be designed to help calibrate it.

Thus we began a new program in {\HST} Cycle 27 (GO-15879) designed to better self-calibrate the parallax offset uncertainty by observing MW Cepheids with photometrically-predicted parallaxes $\pi > 0.8$ mas. In the presence of an additive term (i.e., the poorly known parallax offset) and of a multiplicative term (the Cepheid magnitude scale, corresponding to a constant offset in distance modulus), the combination of the new large-parallax with the prior smaller-parallax set will better break the degeneracy, much as a wide range of dependent and independent samples allows one to simultaneously fit the slope and intercept of a line.  The program resulted in {\HST} photometry of 25 additional Cepheids; for these, and for our previous 50 Cepheids, we can now use the recently released {\Gaia} EDR3 parallaxes \citep{Gaia-Collaboration:2020}.

The rest of the paper is organized as follows.  In \S2 we present the three-filter spatial-scan photometry of the 25 new large-parallax MW Cepheids, together with the previously observed 50, for a total sample of 75 Cepheids with {\HST} photometry, and compare them to ground-based measurements in corresponding passbands.  In \S3 we carry out an analysis of the recently-released {\Gaia} EDR3 parallaxes for all targets; using the precise and accurate {\HST} photometry to recalibrate the extragalactic distance ladder and refine the measurement of the Hubble constant in \S4.
    
\section{Additional Milky Way Cepheid Standards}

R18a,b described the steps used to measure the photometry of bright MW Cepheids from their rapid spatial scans and comparison to ground-based results in similar passbands; we direct interested readers to those publications for details. Photometric measurement uncertainties at a given phase are less than 0.01 mag. R18b compared phase-corrected observations of individual Cepheids across multiple epochs and found average uncertainties in the resulting light-curve mean magnitudes of 0.021, 0.018, and 0.015 mag in {\it F555W}, {\it F814W}, and {\it F160W}, respectively, with the dominant term being a typical $\sim$ 0.02 mag uncertainty in the phase correction to mean light of an individual measurement.

The only changes for this new higher-parallax set of measurements is a 4\% increase in requested scan speed, to $7\farcs 8 / {\rm sec} $, the highest speed available, and the addition of a narrow-band filter with the same effective wavelength as {\it F160W}, {\it F153M}, to check the reliability of photometry for the brightest Cepheids.  We found a zeropoint difference $F160W-F153M$ for the Cepheids of 1.51 mag, in good agreement with the 1.49 magnitude difference 
expected from the STScI calibration and with no significant trend in the difference as a function of flux across 2.5 magnitudes ($0.0036\pm 0.0037$~mag per mag). Phase corrections to mean light are calculated following the procedures outlined in R18a,b, utilizing exclusively $V$ and $I$-band light curves from the same literature sources given in Table 2 of R18b, listed again in the Appendix. \citet{Anderson:2018} estimate that the effect of wide binaries on the photometric calibration of Cepheids is negligible due to the dominance of Cepheids over their companions in flux but will be addressed for any possible astrometric impact on parallaxes in the next section.

For distance measurements and the determination of H$_0$, it is useful to combine the three aforementioned bands into the same reddening-free Wesenheit index \citep{madore82} used by R16 for extragalactic Cepheids in the hosts of SNe~Ia: 
\bq m^W_H=m_{F160W}-0.386(m_{F555W}-m_{F814W}). \eq
\noindent It is useful to compare the {\HST}-system photometry to similar measurements from the ground as done in R18b to test for consistency and to derive useful transformations. For this comparison we make use of the extensive ground-based Cepheid catalog from \cite{Groenwegen:2018} with $V,J,H$ photometry and compare these to the full set of {\HST} Cepheids. We first transform the NIR magnitudes derived from various ground-systems (e.g., SAAO, BIRCAM, CIT) to the 2MASS system following the color transformations given in \cite{Breuval:2020}. These transformations have a mean of 0.01-0.02 mag. We then apply the transformations between the {\HST} and ground systems given below and compare the two in Figure 1. We note the Cepheid CR-Car is not included in the ground-based catalog. The overall agreement is good, as expected since these transformations were derived from a comparison of the {\it HST} Cepheid photometry and that in \cite{Groenwegen:2018}. After identifying a few indicated outliers (assuming fixed errors from the ground catalogs) we derive new transformations of:
\begin{eqnarray}
F555W & = & V+0.202 (J\!-\!H)+0.060 \nonumber \\
F814W & = & V-0.480 (J\!-\!H)-0.025 \nonumber\\
F160W & = & H+0.257 (J\!-\!H)-0.022 \nonumber
\end{eqnarray}
\noindent with dispersions of 0.031, 0.039 and 0.050 mag, respectively, between the {\HST} and ground-transformed magnitudes.  These transformations agree to within 0.01 mag of those given in \citet{Breuval:2020} based on the data in R18b. We note that the mean $J\!-\!H$ color of the Cepheids is 0.43~mag, so there is a substantial zeropoint difference between $H$ and {\it F160W} of $\sim$ 0.08 mag. The ground-to-HST transformation for {\it F160W} includes pre-correcting the {\HST} magnitudes for count-rate non-linearity (hereafter, CRNL) between their fluxes and the brightness of the standard star P330E, a mean correction of 0.017 mag. This ensures the agreement of ground and {\HST} photometry where the {\HST} zeropoints are defined\footnote{The use of ground-based NIR magnitudes with these transformations still requires the addition of 0.030 mag to account for the CRNL that applies to the faint extragalactic Cepheids with HST photometry in R16.}.

Although there are several hundred MW Cepheids observed from the ground for which these transformations can be used, the great value of this sample of 75 is in their photometric consistency. By measuring all Cepheids along the distance ladder (and in both hemispheres) with a single, stable photometric system ({\HST}/WFC3) we can largely eliminate the propagation of zeropoint and bandpass uncertainties among Cepheid flux measurements. As discussed in \S3.2, the lower precision per object means that the weight of the remaining ground sample with good EDR3 parallaxes provides only modest gains.

These 75 $m^W_H$ values have a mean uncertainty of 0.021 mag, including photometric measurement errors, phase corrections, and error propagation to the Wesenheit index, corresponding to approximately 1\% in distance; at the mean expected parallax of 400\ $\mu$as this represents a mean uncertainty of $ 4 \,\mu$as in the {\it predicted} parallax. At this level of precision, both the breadth of the instability strip at 0.04--0.08 mag in $m^W_H$ as seen by \citet{persson04,Macri:2015} and \citet[][hereafter R19]{Riess:2019}, and the expected parallax uncertainties by the end of the {\Gaia} mission (5--$14\,\mu$as) will still dominate the determination of individual Cepheid luminosities. Some of these Cepheids have been suggested as possible binaries; in general, we do not automatically exclude such objects from consideration, rather we use the {\it Gaia} goodness of fit parameter to indicate whether their parallaxes have been compromised by an astrometic binary.
    
In Table 1 we provide the photometric measurements\footnote{$m^W_H$ measurements include a WFC3-IR CRNL correction to account for the 6.4 dex flux ratio in {\it F160W} between these MW Cepheids and the sky-dominated extragalactic Cepheids (R18b).} of these 75 Cepheids for WFC3 {\it F555W}, {\it F814W}, {\it F160W} and $m^W_H$. We also include individual metallicity measurements as compiled by \cite{Groenwegen:2018} for use in the Cepheid \PL relation.

\startlongtable
\begin{deluxetable*}{ccccccccccccccc}
\tabletypesize{\scriptsize}
\tablewidth{0pt}
\tablenum{1}
\tablecaption{Photometric Data for MW Cepheids\label{tb:phot}}
\tablehead{\colhead{Cepheid} &  \colhead{log P} & \colhead{$F555W$} & \colhead{$\sigma$}  & \colhead{$F814W$} & \colhead{$\sigma$}  & \colhead{$F160W^a$} & \colhead{$\sigma$} & \colhead{$m^{W,b}_H$} & \colhead{$\sigma$}  & \colhead{$[Fe/H]^e$} & \colhead{$\pi_{R16}^c$} & \colhead{$\sigma$} & \colhead{$\pi_{EDR3}^d$} & \colhead{$\sigma$}}
\startdata
\hline
\multicolumn{1}{l}{Cycle 22 set} \\
\hline
AA-GEM &  1.053  &  9.9130  &  0.029  &  8.542  &  0.025  &  7.348  &  0.017  &  6.860  &  0.023   &  -0.080  &  0.259  &  0.008  &  0.311  &  0.019  \\
AD-PUP &  1.133  &  10.015  &  0.028  &  8.675  &  0.023  &  7.488  &  0.020  &  7.011  &  0.024   &  -0.060  &  0.214  &  0.006  &  0.254  &  0.018  \\
AQ-CAR &  0.990  &  8.9836  &  0.020  &  7.854  &  0.009  &  6.766  &  0.007  &  6.373  &  0.011   &  0.013  &  0.354  &  0.010  &  0.361  &  0.017  \\
AQ-PUP &  1.479  &  8.8671  &  0.018  &  7.120  &  0.014  &  5.487  &  0.013  &  4.859  &  0.016   &  0.060  &  0.340  &  0.010  &  0.294  &  0.025  \\
BK-AUR &  0.903  &  9.5609  &  0.036  &  8.220  &  0.038  &  7.015  &  0.021  &  6.539  &  0.029   &  0.070  &  0.371  &  0.011  &  0.426  &  0.016  \\
BN-PUP &  1.136  &  10.051  &  0.033  &  8.505  &  0.017  &  7.198  &  0.015  &  6.642  &  0.021   &  0.030  &  0.251  &  0.007  &  0.301  &  0.016  \\
CD-CYG &  1.232  &  9.1207  &  0.011  &  7.468  &  0.012  &  5.900  &  0.012  &  5.307  &  0.014   &  0.120  &  0.398  &  0.011  &  0.394  &  0.018  \\
CP-CEP &  1.252  &  10.757  &  0.015  &  8.638  &  0.052  &  6.871  &  0.022  &  6.095  &  0.030   &  0.050  &  0.270  &  0.008  &  0.279  &  0.022  \\
CR-CAR &  0.989  &  11.750  &  0.019  &  9.973  &  0.018  &  8.384  &  0.014  &  7.736  &  0.017   &  -0.080  &  0.190  &  0.005  &  0.194  &  0.016  \\
CY-AUR$^*$ &  1.141  &  12.052  &  0.012  &  9.953  &  0.020  &  8.106  &  0.025  &  7.334  &  0.027   &  -0.150  &  0.183  &  0.006  &  \nd  &  \nd  \\
DD-CAS &  0.992  &  10.036  &  0.007  &  8.523  &  0.011  &  7.108  &  0.012  &  6.566  &  0.013   &  0.160  &  0.319  &  0.009  &  0.346  &  0.014  \\
DL-CAS$^*$ &  0.903  &  9.1059  &  0.019  &  7.569  &  0.022  &  6.238  &  0.018  &  5.689  &  0.021   &  0.050  &  0.550  &  0.016  &  \nd  &  \nd \\
DR-VEL &  1.049  &  9.7083  &  0.034  &  7.770  &  0.020  &  6.183  &  0.021  &  5.479  &  0.026   &  0.024  &  0.488  &  0.015  &  0.520  &  0.015  \\
GQ-ORI &  0.935  &  8.7199  &  0.020  &  7.632  &  0.024  &  6.523  &  0.032  &  6.146  &  0.034   &  0.250  &  0.418  &  0.013  &  0.408  &  0.023  \\
HW-CAR &  0.964  &  9.2782  &  0.016  &  8.007  &  0.013  &  6.798  &  0.005  &  6.350  &  0.009   &  0.060  &  0.370  &  0.010  &  0.397  &  0.013  \\
KK-CEN &  1.086  &  11.598  &  0.017  &  9.862  &  0.021  &  8.292  &  0.015  &  7.660  &  0.018   &  0.210  &  0.167  &  0.005  &  0.152  &  0.017  \\
KN-CEN &  1.532  &  10.062  &  0.023  &  7.924  &  0.017  &  5.856  &  0.006  &  5.076  &  0.013   &  0.550  &  0.273  &  0.008  &  0.251  &  0.020  \\
RW-CAM$^*$ &  1.215  &  8.8673  &  0.015  &  7.044  &  0.014  &  5.451  &  0.021  &  4.794  &  0.022   &  0.080  &  0.519  &  0.015  &  \nd  &  \nd  \\
RW-CAS &  1.170  &  9.3719  &  0.021  &  7.863  &  0.016  &  6.483  &  0.022  &  5.944  &  0.024   &  0.280  &  0.322  &  0.010  &  0.334  &  0.021  \\
RY-CAS &  1.084  &  10.075  &  0.019  &  8.333  &  0.040  &  6.715  &  0.010  &  6.085  &  0.020   &  0.320  &  0.342  &  0.010  &  0.359  &  0.016  \\
RY-SCO &  1.308  &  8.2067  &  0.012  &  6.206  &  0.010  &  4.408  &  0.010  &  3.685  &  0.012   &  0.010  &  0.757  &  0.021  &  0.764  &  0.035  \\
RY-VEL &  1.449  &  8.5234  &  0.036  &  6.757  &  0.016  &  5.211  &  0.017  &  4.576  &  0.023   &  0.090  &  0.403  &  0.012  &  0.376  &  0.023  \\
S-NOR &  0.989  &  6.5779  &  0.011  &  5.410  &  0.012  &  4.391  &  0.012  &  3.990  &  0.014   &  0.100  &  1.054  &  0.030  &  1.099  &  0.024  \\
S-VUL$^*$$^*$ &  1.839  &  9.1668  &  0.008  &  6.862  &  0.012  &  4.885  &  0.010  &  4.043  &  0.011   &  0.090  &  0.287  &  0.008  &  0.237  &  0.022  \\
SS-CMA &  1.092  &  10.121  &  0.012  &  8.444  &  0.008  &  6.894  &  0.011  &  6.289  &  0.012   &  0.012  &  0.315  &  0.009  &  0.308  &  0.014  \\
SV-PER$^*$ &  1.046  &  9.2186  &  0.016  &  7.760  &  0.014  &  6.435  &  0.027  &  5.916  &  0.028   &  0.030  &  0.400  &  0.012  &  \nd  &  \nd  \\
SV-VEL &  1.149  &  8.7316  &  0.026  &  7.302  &  0.009  &  6.024  &  0.010  &  5.517  &  0.015   &  0.090  &  0.411  &  0.012  &  0.434  &  0.019  \\
SV-VUL$^*$$^*$ &  1.653  &  7.2675  &  0.047  &  5.648  &  0.033  &  4.214  &  0.027  &  3.639  &  0.035   &  0.110  &  0.457  &  0.015  &  0.402  &  0.023  \\
SY-NOR$^*$ &  1.102  &  9.8284  &  0.023  &  7.925  &  0.038  &  6.214  &  0.013  &  5.523  &  0.022   &  0.230  &  0.435  &  0.013  &  \nd  &  \nd  \\
SZ-CYG &  1.179  &  9.6209  &  0.013  &  7.756  &  0.017  &  6.004  &  0.008  &  5.329  &  0.012   &  0.150  &  0.426  &  0.012  &  0.445  &  0.014  \\
T-MON &  1.432  &  6.0680  &  0.023  &  4.828  &  0.016  &  3.725  &  0.021  &  3.298  &  0.024   &  0.040  &  0.749  &  0.022  &  0.745  &  0.057  \\
U-CAR &  1.589  &  6.3852  &  0.038  &  4.967  &  0.023  &  3.768  &  0.019  &  3.272  &  0.026   &  0.250  &  0.589  &  0.018  &  0.561  &  0.025  \\
UU-MUS &  1.066  &  9.9212  &  0.024  &  8.457  &  0.025  &  7.108  &  0.010  &  6.584  &  0.017   &  0.190  &  0.282  &  0.008  &  0.306  &  0.013  \\
V-339-CEN &  0.976  &  8.8402  &  0.024  &  7.321  &  0.016  &  5.990  &  0.024  &  5.448  &  0.026   &  -0.080  &  0.557  &  0.017  &  0.568  &  0.023  \\
V-340-ARA &  1.318  &  10.460  &  0.024  &  8.554  &  0.014  &  6.808  &  0.012  &  6.115  &  0.016   &  -0.080  &  0.245  &  0.007  &  0.239  &  0.022  \\
VW-CEN &  1.177  &  10.379  &  0.031  &  8.718  &  0.023  &  7.158  &  0.010  &  6.558  &  0.018   &  0.410  &  0.238  &  0.007  &  0.260  &  0.017  \\
VX-PER &  1.037  &  9.4589  &  0.008  &  7.906  &  0.006  &  6.470  &  0.009  &  5.914  &  0.010   &  0.030  &  0.407  &  0.011  &  0.392  &  0.019  \\
VY-CAR &  1.276  &  7.6162  &  0.014  &  6.253  &  0.007  &  4.991  &  0.004  &  4.513  &  0.007   &  0.080  &  0.539  &  0.015  &  0.565  &  0.018  \\
VZ-PUP &  1.365  &  9.7715  &  0.033  &  8.262  &  0.022  &  6.931  &  0.017  &  6.390  &  0.023   &  -0.010  &  0.200  &  0.006  &  0.220  &  0.016  \\
WX-PUP &  0.951  &  9.1909  &  0.030  &  7.944  &  0.012  &  6.807  &  0.010  &  6.368  &  0.016   &  -0.010  &  0.376  &  0.011  &  0.387  &  0.017  \\
WZ-SGR &  1.339  &  8.2021  &  0.012  &  6.481  &  0.013  &  4.858  &  0.009  &  4.242  &  0.011   &  0.280  &  0.547  &  0.015  &  0.612  &  0.031  \\
X-CYG &  1.214  &  6.5295  &  0.020  &  5.230  &  0.049  &  4.080  &  0.033  &  3.629  &  0.039   &  0.160  &  0.883  &  0.029  &  0.910  &  0.022  \\
X-PUP &  1.414  &  8.6949  &  0.019  &  7.128  &  0.010  &  5.628  &  0.008  &  5.069  &  0.012   &  0.020  &  0.341  &  0.010  &  0.397  &  0.022  \\
XX-CAR &  1.196  &  9.4627  &  0.027  &  8.067  &  0.015  &  6.833  &  0.022  &  6.337  &  0.025   &  0.010  &  0.264  &  0.008  &  0.305  &  0.016  \\
XY-CAR &  1.095  &  9.4660  &  0.011  &  7.927  &  0.009  &  6.455  &  0.006  &  5.904  &  0.008   &  0.012  &  0.375  &  0.010  &  0.390  &  0.015  \\
XZ-CAR &  1.221  &  8.7725  &  0.017  &  7.217  &  0.006  &  5.770  &  0.007  &  5.215  &  0.010   &  0.026  &  0.425  &  0.012  &  0.473  &  0.020  \\
YZ-CAR &  1.259  &  8.8644  &  0.016  &  7.401  &  0.007  &  5.991  &  0.013  &  5.471  &  0.015   &  -0.030  &  0.359  &  0.010  &  0.358  &  0.020  \\
YZ-SGR &  0.980  &  7.4662  &  0.021  &  6.176  &  0.014  &  5.103  &  0.020  &  4.653  &  0.022   &  0.120  &  0.786  &  0.023  &  0.860  &  0.027  \\
Z-LAC &  1.037  &  8.5686  &  0.022  &  7.157  &  0.015  &  5.917  &  0.018  &  5.417  &  0.021   &  0.070  &  0.509  &  0.015  &  0.510  &  0.023  \\
\hline
\multicolumn{1}{l}{Cycle 27 set} \\
\hline
AG-CRU &  0.584  &  8.3175  &  0.013  &  7.307  &  0.011  &  6.414  &  0.027  &  6.068  &  0.028   &  0.020  &  0.748  &  0.023  &  0.758  &  0.022  \\
AP-PUP &  0.706  &  7.4560  &  0.016  &  6.412  &  0.014  &  5.534  &  0.027  &  5.177  &  0.028   &  -0.020  &  0.941  &  0.029  &  0.924  &  0.022  \\
AP-SGR &  0.704  &  7.1056  &  0.028  &  6.036  &  0.013  &  5.094  &  0.027  &  4.729  &  0.030   &  0.160  &  1.145  &  0.035  &  1.217  &  0.026  \\
BF-OPH &  0.609  &  7.5091  &  0.018  &  6.347  &  0.010  &  5.374  &  0.027  &  4.972  &  0.028   &  0.110  &  1.184  &  0.036  &  1.189  &  0.026  \\
BG-VEL &  0.840  &  7.7827  &  0.010  &  6.299  &  0.009  &  5.054  &  0.019  &  4.529  &  0.020   &  0.040  &  1.033  &  0.030  &  1.045  &  0.019  \\
ER-CAR &  0.888  &  6.9095  &  0.011  &  5.916  &  0.012  &  5.078  &  0.027  &  4.742  &  0.028   &  0.120  &  0.867  &  0.026  &  0.869  &  0.016  \\
R-CRU &  0.765  &  6.8479  &  0.017  &  5.856  &  0.016  &  4.984  &  0.027  &  4.649  &  0.028   &  0.100  &  1.088  &  0.033  &  1.078  &  0.031  \\
R-MUS &  0.876  &  6.4568  &  0.009  &  5.447  &  0.008  &  4.609  &  0.019  &  4.268  &  0.020   &  -0.110  &  1.117  &  0.033  &  1.076  &  0.019  \\
R-TRA &  0.530  &  6.7236  &  0.013  &  5.794  &  0.014  &  5.025  &  0.019  &  4.714  &  0.020   &  0.160  &  1.497  &  0.044  &  1.560  &  0.018  \\
RV-SCO &  0.783  &  7.1616  &  0.010  &  5.871  &  0.007  &  4.773  &  0.019  &  4.323  &  0.020   &  0.080  &  1.234  &  0.036  &  1.257  &  0.023  \\
RX-CAM$^*$ &  0.898  &  7.8310  &  0.016  &  6.215  &  0.013  &  4.791  &  0.028  &  4.216  &  0.029   &  0.080  &  1.090  &  0.034  &  \nd  &  \nd  \\
RY-CMA &  0.670  &  8.2358  &  0.015  &  7.111  &  0.013  &  6.045  &  0.027  &  5.656  &  0.028   &  0.140  &  0.787  &  0.024  &  0.825  &  0.032  \\
S-CRU$^e$ &  0.671  &  6.6700  &  0.050  &  5.698  &  0.011  &  4.843  &  0.027  &  4.516  &  0.033   &  0.080  &  1.335  &  0.042  &  1.342  &  0.026  \\
S-TRA &  0.801  &  6.5171  &  0.013  &  5.553  &  0.012  &  4.752  &  0.027  &  4.429  &  0.028   &  0.010  &  1.150  &  0.035  &  1.120  &  0.024  \\
SS-SCT &  0.565  &  8.3122  &  0.010  &  7.073  &  0.005  &  6.034  &  0.019  &  5.600  &  0.019   &  0.110  &  0.948  &  0.028  &  0.934  &  0.025  \\
T-VEL &  0.667  &  8.1205  &  0.009  &  6.915  &  0.007  &  5.839  &  0.019  &  5.419  &  0.020   &  -0.160  &  0.904  &  0.026  &  0.940  &  0.018  \\
TX-CYG &  1.168  &  9.6108  &  0.024  &  7.083  &  0.015  &  4.789  &  0.027  &  3.862  &  0.029   &  0.260  &  0.844  &  0.026  &  0.829  &  0.020  \\
U-AQL$^*$ &  0.847  &  6.5396  &  0.019  &  5.168  &  0.029  &  4.115  &  0.027  &  3.636  &  0.030   &  0.140  &  1.531  &  0.047  &  \nd  &  \nd  \\
U-SGR &  0.829  &  6.8864  &  0.018  &  5.388  &  0.011  &  4.143  &  0.027  &  3.615  &  0.028   &  0.140  &  1.588  &  0.049  &  1.605  &  0.025  \\
V-CAR &  0.826  &  7.4753  &  0.009  &  6.403  &  0.008  &  5.463  &  0.019  &  5.096  &  0.020   &  0.080  &  0.810  &  0.024  &  0.797  &  0.015  \\
V-VEL &  0.641  &  7.5198  &  0.013  &  6.555  &  0.010  &  5.693  &  0.027  &  5.366  &  0.028   &      0  &  0.951  &  0.029  &  0.953  &  0.019  \\
V0386-CYG &  0.721  &  9.8126  &  0.015  &  7.748  &  0.014  &  5.944  &  0.027  &  5.192  &  0.028   &  0.170  &  0.901  &  0.028  &  0.894  &  0.014  \\
V0482-SCO &  0.656  &  8.0697  &  0.013  &  6.773  &  0.013  &  5.697  &  0.027  &  5.242  &  0.028   &  0.019  &  0.982  &  0.030  &  0.993  &  0.028  \\
V0636-SCO &  0.832  &  6.8167  &  0.009  &  5.618  &  0.008  &  4.568  &  0.020  &  4.154  &  0.021   &  0.070  &  1.239  &  0.036  &  1.180  &  0.037  \\
W-GEM &  0.898  &  7.0841  &  0.057  &  5.899  &  0.018  &  4.863  &  0.027  &  4.454  &  0.036   &  -0.010  &  0.984  &  0.032  &  1.006  &  0.031  \\
\hline
\enddata
\tablecomments{$^a$Does not include addition of $0.0075 \pm 0.006 $ mag/dex to correct CRNL for 5 to 6.5 dex between MW and extragalactic Cepheids.}
\tablecomments{$^b$Includes addition of CRNL to allow direct comparison to extragalactic Cepheids in R16 which lack any CRNL correction.}
\tablecomments{$^c$ $\pi_{phot}=10^{-0.2(\mu-10)}$ where $\mu=m^W_H-M^W_H$, and $M^W_H$ is the absolute Wesenheit magnitude determined from the Cepheid period and the distance scale from \cite{Riess:2016} where $b_W\!=\!-3.26$, $Z_W\!=\!-0.17$ mag/dex, $M^W_{H,1}\!=\!-5.93$ mag which results in $H_0\!=\!73.24$\kmsmpc as discussed in the text.}
\tablecomments{$^d$ Includes L20b parallax offset, does not include addition of best-fit residual parallax offset found here, -14\muas. EDR3 errors increased by 10\%.}
\tablecomments{$^e$ S Cru in $F555W$ transformed from $v$ in \citet{Groenwegen:2018} due to HST failed acquisition.}
\tablecomments{$^*$ Unreliable EDR3 parallax, see text.}
\tablecomments{$^{**}$ Possible outlier, see text.}
\end{deluxetable*}

\section{{\Gaia} EDR3}

The quality of the parallaxes of MW Cepheids has markedly improved from {\Gaia} DR2 to EDR3, as we will see in the rest of this Section.  The improvements result from an increase in the sampling (34 vs.~22 months), improved analysis of the data \citep[][hereafter L20a,b]{Lindegren:2020a,Lindegren:2020b}, and an improved characterization of the leading systematic uncertainty in DR2: the parallax offset term.  We present a more detailed discussion of the changes in EDR3, and how they affect the quality of Cepheid parallaxes, in the Discussion; here we proceed with the analysis of the EDR3 parallaxes with the recommended parallax offset (L20a, b).

We use the formulation of L20b to calibrate the parallaxes, using parallax offsets that are a function of source color ($ \nu_{\rm eff} $ or pseudocolor), $G$-band magnitude, and ecliptic latitude $\beta$. Our {\HST} sample of 75 MW Cepheids occupies a modest range of color space as expected for stars with F-K spectral types: median $ \nu_{\rm eff} $ of 1.42~$\mu$m$^{-1}$ with a dispersion of 0.055~$\mu$m$^{-1}$  and a full range of 1.30 to 1.53~$\mu$m$^{-1}$, corresponding to a median {\it F555W}$-${\it F814W} of 1.46~mag and a full range of 0.93 to 2.53~mag. The median $G$ is 8.3~mag with a dispersion of 1.3~mag and a full range of 6.1 to 11.2~mag. The Cepheids are well distributed in ecliptic latitude, $-72^\circ\!<\!\beta\!<\!+62^\circ$. Among the three properties that determine the EDR3 parallax offset following the L20b prescription, ecliptic latitude dominates the variation in this value for our sample, as shown in Figure 2. The dependence follows a parabolic function of $\beta$, reaching a minimum of $-38$\mias near $\beta=0$ and increasing to $-20$\mias at $|\beta|\approx 50^\circ$. The median parallax offset for the Cepheid sample is $-24$\mias with a dispersion of 9\mias and a full range of $-38$ to $-4$\muas. The dispersion perpendicular to the parabolic dependence on $\beta$ drops to 1.8\muas. One Cepheid (CY Aur, the faintest in our sample) diverges from the $\beta$ dependence. Its mean magnitude is near a sharp inflection point in the L20b formulae at $G=11$~mag, where the parallax offset changes from $-43$ to $-24$\mias between $G\!=\!10.8$ and 11.2~mag. Given this large range of offset values and since the brightness of CY Aur may vary across this boundary, we will cautiously treat its parallax as unreliable.

It is reasonable to expect some residual uncertainty in the parallax offset in the small magnitude and color range of these Cepheids. L20b suggests an uncertainty of ``a few microarcseconds'' in the parallax offset across the well-calibrated range. Because our Cepheids are at the bright end of this range, we will adopt a somewhat more conservative {\it a priori} uncertainty of 10 {\muas} for the L20b parallax offset. The \PL relation itself provides a strong tool to refine the offset in this range, as we will show in the following.

Not all Cepheids can be expected to yield useful parallaxes from EDR3. The most likely reason for a bad parallax is binarity with a period close to one year or a close association with a PSF that blends with the Cepheid. L20a recommends the use of the goodness of fit (GOF) to identify compromised parallaxes. Two of our Cepheids, RW Cam and SV Per, have GOF$>\!100$ and were seen by R18b with {\HST} imaging to have companions blended within $0\farcs 2$. There are four others with high GOFs of 18 to 28, all in known binaries: U Aql \citep{Gallenne:2019}, DL Cas \citep{Evans:1995}, SY Nor \citep{Kervella:2019} and AD Pup \citep{Szabados:2013}. The rest of the sample has GOF values of 12.5 or lower with no major gaps, so we set this as the threshold for inclusion and will later check for outliers. This leaves us with 68 Cepheids, as indicated in Table 1.

\subsection{Photometric and astrometric parallaxes}

We compare the EDR3 Cepheid parallaxes to their photometrically-predicted values using the Cepheid \PL relation used to measure H$_0$ (R16). It is advantageous to work in ``parallax space'' to retain the Gaussian description of the {\Gaia} EDR3 parallax errors.

From the definition of the Wesenheit index in Eq.~1, the {\it photometric} distance modulus of a Cepheid is given by the difference in magnitudes of an apparent and absolute flux, $\mu_0=m_H^W-M_H^W$. This is expressed following the \PL relation in R16 for the $i$th Cepheid as \bq \mu_{0,i}=m_{i,H}^W-(M_{H,1}^W+b_W\, (\log\,P_{i}-1)+Z_W \ \Delta {\rm [O/H]}_{i}), \label{eq:cephmagalt} \eq

\noindent where $M_{H,1}^W$ is the absolute magnitude\footnote{$M_{H,1}^W$ was defined at $P=1$~d in R16, and was changed here to $\log P=1$ ($P=10$~d) for consistency with \citet{Breuval:2020}.} for a Cepheid with $\log P=1$ ($P=10$~d) and solar metallicity, while $b_W$ and $Z_W$ define the relation between Cepheid period, metallicity, and luminosity. The apparent magnitude, $m_{H}^W$ is given in Eq.~1. The distance modulus is $\mu_0=5 \log D + 25$, with $D$ the luminosity distance in Mpc.

The expected parallax, $\pi_{phot,i}$ in units of mas, is given by \bq \pi_{phot,i}=10^{-0.2(\mu_{0,i}-10)} \eq

With negligible uncertainties in the periods, the mean uncertainties in the predicted parallaxes are 1\% in distance from the photometric measurements of the previous section and $\sim\!2-3$\% in distance due to the width of the instability strip. Because the photometry uncertainties are very small, they are very close to symmetric in parallax or distance, to better than a tenth of a percent.

In Figure 3 we compare the parallaxes using the values of $b_W\!=\!-3.26$, $M_{H,1}^W\!=\!-5.93$~mag, and $Z_W\!=\!-0.17$~mag/dex from R19. With {\it no free parameters} and no additional characterization of the parallax offset, the agreement between parallaxes appears quite good. The improvement from the comparable result with DR2 and the smaller {\HST} sample from R18b, also shown in Figure 3, is {\it striking}. The improved precision of the parallaxes is evident and most easily seen at lower-parallax values where the {\HST} sample is unchanged.

Closer scrutiny of the residuals indicates a modest {\it over-correction} of the parallax offset, with a median difference of 15 \mias (formally significant at $\sim$ $3\sigma$) and no visible correlation at the 1-$\sigma$ level with parallax, either measured or predicted (which, if present, would indicate a distance scale term). There are also two Cepheids (S Vul and SV Vul) near the boundary of Chauvenet's outlier criterion for a sample of this size ($\sim\!2.6\sigma$); we  tentatively exclude these two objects, but we will give our final results with and without them. We also show in Figure 3 the predictions of the Cepheid parallaxes for a range of values of H$_0$. 

Our goal, as in R18b, is to simultaneously determine two parameters: the optimal parallax offset applicable to bright Cepheids, an additive term to parallax, and the calibration of the distance scale, a multiplicative term of parallax. However, rather than assuming the other parameters which characterize the \PL relations of MW and extragalactic Cepheids are the same, we first undertake a more general 4-parameter analysis including the slope $b_W$ and metallicity term $Z_W$ defined above, to determine their consistency.

Therefore we seek to optimize the value of: \bq \chi^2=\sum {( \pi_{{\rm EDR3},i} - \pi_{{\rm phot},i} + {\it zp})^2 \over \sigma_i^2}, \eq
\noindent where {\it zp} is a residual parallax offset {\it after} application of the L20b-derived parallax offset and $\pi_{phot,i}$ is a function of the Cepheid \PL parameters $b_W, M_{H,1}^W, Z_W$ (as in given in equations 2 and 3). Note that these parameters are separable as {\it zp} is additive to the photometric parallaxes, $M_{H,1}^W$ is multiplicative, and $b_W, Z_W$ depend on individual periods and metallicities.

We determine the individual $\sigma_i$ by adding in quadrature the photometric parallax uncertainty, the intrinsic width of the NIR Wesenheit $P-L$ (0.06 mag)  and the parallax uncertainty given in the EDR3 release. Based on some suggestion of possible excess uncertainty in the {\Gaia} EDR3 data validation \citep[see Fig.~21 of][]{Fabricius:2020}, we conservatively increase the nominal parallax uncertainty assigned in the EDR3 release by 10\% (an augmentation by 30\% was indicated for the prior DR2 as discussed in L18b and R18b, so excess uncertainty appears to be less for EDR3). The mean of the EDR3 uncertainties is 21\mias (median 20\muas), while the mean of the full $\sigma_i$ is 29\mias (median 27\muas). 

Minimizing the value of $\chi^2$ gives values of $b_W$=\fourslope, $Z_W$=\fourmetal mag dex$^{-1}$, and {\it zp}$=$ \fourzp \mias, with a value of $\chi^2=$ {\fourchi} for 66 degrees of freedom. The values of $b_W$ and $Z_W$ found here for the MW Cepheids are found to be fully consistent with the extragalactic Cepheids in R16 and R19, though they are determined with much lower precision here. The value of $M_{H,1}^W$ is not readily applicable to other Cepheids on the distance ladder because it is not determined for the same \PL relation, i.e., one using the same values of $b_W$ and $Z_W$ as in R16 or R19, which we remedy below.

We now calibrate the luminosity of Cepheids along the distance ladder by adopting fixed values of $b_W=-3.26$ and $Z_W=-0.17$ mag dex$^{-1}$ for the slope and the metallicity term, these are the values derived by R19 from other galaxies (i.e., the LMC, M31, NGC$\,$4258, and 19 SN Ia hosts\footnote{Future analyses would ideally optimize the value of $b_W$ and $Z_W$ across all Cepheid hosts, but there is little difference in practice, as these parameters are far better determined from the aforementioned extragalactic samples of R16 and R19.}) and again optimize the value of $\chi^2$ for the two free parameters, {\it zp} and $M_{H,1}^W$. This is the same procedure used in R18b.

We find {\it zp}$=$\twozp \mias and $M_{H,1}^W=${\twomag} mag with $\chi^2=${\twochi} for 66 degrees of freedom. Applied to the distance ladder from R16 and R19 to calibrate SNe Ia yields H$_0=${\hoedr} \kmsmpc. Confidence regions for the two parameters are shown in Figure 4. Although these two parameters are correlated, the range of Cepheid parallaxes ($0.2-1.5$\mas) largely breaks their degeneracy to provide a calibration of the distance ladder with 1\% precision, better than any other individual geometric calibration (R19). There is no evidence of a correlation of the residuals with $G$ mag ($0.4\sigma$), color ($1.0\sigma$) or ecliptic latitude ($0.2\sigma$), the 3 parameters used by L20b to characterize the parallax offset. We also note that the mean metallicity of the {\HST} sample is 0.09 dex \citep{Groenwegen:2018}, slightly greater than solar, and with the empirical metallicity term, these MW Cepheids are expected to be 0.015 mag brighter on average than the definition of $M_{H,1}^W$.  Including the two marginal outliers discussed in the prior section (S Vul and SV Vul) yields {\it zp}$=$\twozpwss \mias and $M_{H,1}^W=$\twomagwss\ mag with $\chi^2=$\twochiwss, similar parameters but with the expected higher $\chi^2$ for 68 degrees of freedom.

If we do not include freedom for {\it zp} but rather adopt the exact L20b value, we find $M_{H,1}^W=${\onemag} mag and H$_0=${\hoedrone} \kmsmpc, implying a calibration with  0.6\% precision. However, this solution has a significantly greater $\chi^2=${\onechi} and ignores the strong evidence for a residual parallax offset, detected here at $2.5\sigma$ with the assumed 10 \mias uncertainty in the fiducial L20b parallax offset, or more realistically $3\sigma$ ($-17\pm 6 $\muas) with no prior on the quality of the L20b offset.

The size and direction of a residual parallax offset is also corroborated by the analysis of red giants with asteroseismic data from Kepler \citep{Zinn:2018}. \citet{Zinn:2020} finds an approximate residual parallax offset of $-15 \pm 5$\mias (same sense of an overall smaller parallax offset as found here) for brighter red giants in the range of $G= 10-11$ mag\footnote{\citet{Zinn:2020} finds good agreement with the parallax offset of L20b for red giants fainter than $G=11$}  and colors similar to our Cepheids. If we use this as a prior, rather than the nominal 10~{\muas} uncertainty on the L20b offset, we find a tighter constraint of $M_{H,1}^W$ to {\twomagzinn} mag, a remarkable 0.85\% foundation for determining H$_0$ that points to the room for improvement with {\Gaia} DR4.  In the following, we conservatively adopt the nominally less precise calibration internal to the Cepheids, which effectively marginalizes over the uncertainty in the parallax offset, and without the red giant prior. Future characterization of the EDR3 parallax offset may further justify the use of a tighter constraint.

The value of $M_{H,1}^W$ is quite consistent with the value of $-5.93$~mag from R16, indicating that the predicted parallaxes, after accounting for the offset, are in good agreement with EDR3, and further affirming the cosmic distance scale with the value of H$_0$ used to predict the parallaxes from R16. On the other hand, this value of $M_{H,1}^W$ is inconsistent with the value of $-6.12$~mag, needed to match the Planck CMB$+\Lambda$CDM value of H$_0$, at the \faceplanck $\sigma$ confidence level (99.997\% likelihood), confirming again the ``$H_0$ tension'' (see \citealt{Verde:2019} for a review).

\subsection{ Ground-Based Sample, Caveats}  

We might improve the constraint on {\it zp} and $M_{H,1}^W$ (or even $b_W$ and $Z_W$) by considering a larger sample of MW Cepheids, though the augmentation of the sample would need to rely exclusively on ground-based photometry. Using the \cite{Groenwegen:2018} catalog of $V,J,H,K$ photometry and the photometric transformations in \S2 for fundamental-mode Cepheids would augment the {\HST} sample by an additional $\sim\!200$ Cepheids which have good-quality parallaxes. However, the transformed photometric uncertainty per object increases, matching the width of the instability strip, so that the statistical significance of the additional sample only just matches the {\HST} sample. More concerning is that the {\HST} sample was selected to have low extinction ($A_H< 0.4$ mag for the 50 Cepheids in R18b and $A_H < 0.6$ mag for the additional 25 with larger parallaxes presented here), so the additional Cepheids are mostly highly-reddened and distant, so these Cepheids may offer less precise results. Our trial analyses using the ground-based sample yielded similar parameters as for the {\HST} sample, but with a larger dispersion which may require additional modeling of the uncertainties. It is not clear if the additional dispersion may result from {\Gaia} uncertainties over a different range of measurement space, inhomogeneities between ground surveys, uncertainties in larger reddenings or some combination of these. We therefore chose to focus on the better-understood and more precise {\HST} sample.

\begin{deluxetable}{lllllll}
\tablewidth{0pc}
\tablecaption{Best Fits to Gaia EDR3 \label{tb:h0c}}
\tablenum{2}
\tabletypesize{\small}
\tablewidth{0pc}
\tablehead{\colhead{Fit} & \colhead{$M_{H,1}^W$} & \colhead{{\it zp}} & \colhead{$b_W$} & \colhead{$Z_W$} & \colhead{$\chi^2$} & \colhead {H$_0$}\\
\colhead{} & \colhead {[mag]} & \colhead {[\muas]} & \multicolumn{2}{c}{[mag/dex]} & \colhead{} & \colhead{[\kmsmpc]}}
\startdata
4-parameter & $-5.915\pm 0.030$ & \fourzp & \fourslope & \fourmetal & \fourchi & $(73.0 \pm 1.4^a)$ \\
4-parameter with outliers & $-5.930\pm 0.030$ & \fourzpwss & \fourslopewss & \fourmetalwss & \fourchiwss & $(72.5 \pm 1.4^a)$  \\
\hline
{\bf 2-parameter (best)} & \twomag & \twozp & $-3.26^b$ & $-0.17^b$ & \twochi &\ \hoedr \\
2-parameter with outliers & \twomagwss & \twozpwss & $-3.26^b$ & $-0.17^b$ & \twochiwss &\ \hoedrss \\
\hline
1-parameter & \onemag & \multicolumn{1}{c}{0$^c$} & $-3.26^b$ & $-0.17^b$ & \onechi &\ \hoedrone \\
\hline
\hline
\enddata
\tablecomments{{\it a}: Cepheid luminosity not determined with same \PL parameters $b_W$ and $Z_W$ from R19, so not directly applicable to determine H$_0$. {\it b}: Fixed to R19 values. {\it c}: Assuming no residual parallax offset in {\Gaia} EDR3.}
\end{deluxetable}
  
\section{Discussion}

\subsection{The Parallax Offset in EDR3 Parallaxes}

{\Gaia} EDR3 \citep{Gaia-Collaboration:2020} contains full astrometric single-source solutions and three-band photometry, including their open-filter $G$ magnitude and two-color photometry in the integrated $G_{BP}$ and $G_{RP}$, for nearly 1.5 billion sources. In many respects, the solutions presented in EDR3 are similar to those in DR2 \citep{Gaia-Collaboration:2018}, although the solution covers a longer time period (34 vs. 22 months), and has smaller formal uncertainties. However, the calibration model used in EDR3 is far more extensive, and includes several additional parameters motivated by trends seen in preliminary solutions (\S3.3 of L20a). In addition, the EDR3 solution makes extensive use of color information for each source in order to remove chromaticity effects, especially color-dependent offsets in the PSF position in the {\Gaia} focal place (see \S2.3 of L20a). The use of color information in EDR3 comes in two different flavors. The standard solution, obtained for a majority of the sources, uses the effective wavenumber $ \nu_{\rm eff} $ calculated from the sampled and calibrated spectra in the blue and red photometers, to estimate chromaticity effects in the PSF; these are described as five-parameter solutions and are identified in the EDR3 catalog by having $\texttt{astrometric\_params\_solved} = 31$. However, about 12\% of the sources brighter than $G=18$ (and two-thirds overall) lack a valid $ \nu_{\rm eff} $ for various reasons; for these, the astrometric data themselves were used to estimate the chromatic shift of the PSF in each observation. The resulting solution has six parameters: the five standard astrometric parameters and a ``pseudocolor'', i.e., a color estimate that minimizes the residuals in the astrometric solution. Such solutions are identified in the EDR3 catalog by having $\texttt{astrometric\_params\_solved} = 95$. Most of the 75 parallaxes used in this work come from the five-parameter EDR3 solution, with 12 coming from the six-parameter solution. 

Some limitations remain in the EDR3 solution. One which could be a consideration for our targets is that the solution assumes that the color of a source is the same in all observations. This is a good approximation for most stars, but it does not fully apply to our targets, which change magnitude and color according to their phase. For most bright stars, the effective wavenumber could in principle be determined independently for the majority of the observations (or, in the case of Cepheids with known light curve ephemeris, determined on the basis of the phase for each exposure); however, this step is not yet included in the EDR3 pipeline (\S2.3 of L20a). In practice, the color variation of a Cepheid, about $\pm 0.25$ mag in $ G_{BP}- G_{RP} $ during one cycle, will add a small amount of astrometric noise, and an even smaller amount of parallax noise. Since the parallax measurement is well averaged over all phases such noise will be statistical and well below the present statistical parallax errors.

Second, and more significant, is the parallax offset due to the variation of the basic angle discovered during {\Gaia} commissioning (L18a). Most of the basic angle variation can be constrained as part of the astrometric solution, but as noted by \citet{Butkevich:2017}, a near-degeneracy remains, which manifests itself as a parallax offset error. This parallax offset was determined empirically  for DR2 on the basis of quasar parallax measurements (L18a). However, L18 already noted some hints of a variation of the mean quasar parallax with color, magnitude, and position in the sky, although the quasar sample lacked the breadth in magnitude and color to cover the full span of {\Gaia} DR2 sources. The variation of the parallax offset was confirmed by several subsequent studies (R18b, \cite{Zinn:2018,Arenou:2018}), which found significantly different offsets for different samples of stars, with a strong dependence of the offset, e.g., with magnitude.

The parallax offset has been studied very carefully for EDR3 (L20b). The authors have used a combination of quasar, LMC stars, and physically bound pairs to constrain the parallax offset and its variation over a broad range of magnitude, color, and ecliptic latitude. The parallax for all quasars is assumed to be zero; the parallax for all LMC stars is assumed to be the same, but no assumption is made on its value. For stars in physical pairs, the assumption is that both elements have the same parallax. On the basis of these data, the authors obtained an approximate expression for the mean parallax offset as a function of magnitude $G$, color, and ecliptic latitude. The color parameter is the effective wavelength $ \nu_{\rm eff} $ for five-parameter solutions, and the pseudocolor for six-parameter solutions; separate expressions are derived for the two cases. In principle, subtracting the estimated parallax offset from the catalog parallax should remove the broad dependencies from magnitude, color, and position; as seen in our analysis this approach is largely successful, and leads to a significant improvement of the data with no trend of residuals with magnitude, color, or ecliptic latitude. However, the Quasar data does not meaningfully calibrate the offset at magnitudes brighter than $G=14$ nor the LMC data brighter than $G=11$.  To reach the magnitude range of Cepheids ($G=6-10$) requires inferring the offsets of the fainter star in physical pairs (via the Quasar and LMC-derived formulae) and using the brighter stars in the pairs to infer the offset at lower magnitudes.  This process is repeated with even brighter pairs to extend to $G=6$.  Thus the offset for Cepheids is expected to be less well constrained.  It is therefore not surprising that we do find evidence of a residual offset and the need to marginalize over this term lowers the precision available to calibrate the uncertainty in the Hubble constant from 0.6\% to 1.0\%, a quite important cost in the quest to determine H$_0$ to 1\% precision.

\subsection{The Status of the Hubble Constant}

It is quite reasonable to expect the precision of each of the three steps in the distance ladder linking geometry, Cepheids and SN Ia to measure H$_0$ to be determined to better than 1\% in the very near future.  We expect that the geometric calibration of Cepheids will approach 0.5\% precision by DR4, matching the current precision of the SN Ia Hubble diagram \citep{Scolnic:2018}. With the number of high quality calibrations of SN Ia with Cepheids approaching 40, a total uncertainty  in the range of 1.0\% to 1.3\%  in H$_0$ (depending on residual systematics) appears within reach.   

We expect that {\Gaia} EDR3 will also impact the calibration of other distance indicators. EDR3 puts the parallaxes of globular clusters, most notably {\hbox{$\omega$~Cen}}, in useful range for the first time; as a result, the direct geometric calibration of the Tip of the Red Giant Branch (TRGB) is now within reach (Soltis et al  2020, in prep). 

We may hope that enhanced measurements or theoretical insights will lead to an {\it explanation} of the present $\sim 5 \sigma$ tension between direct determinations of the Hubble constant and values inferred from $\Lambda$CDM calibrated with Early Universe physics and the CMB (see \citealp{Verde:2019} for review). 

\bigskip

\bigskip

\acknowledgements

We are grateful to the entire {\Gaia} collaboration for providing data and assistance which made this project possible. We congratulate them on their tremendous achievement to date. We acknowledge with thanks the variable star observations from the AAVSO International Database contributed by observers worldwide and used in this research.

Support for this work was provided by the National Aeronautics and Space Administration (NASA) through programs GO-12879, 13334, 13335, 13344, 13571, 13678, 13686, 13928, 13929, 14062, 14394, 14648, 14868, 15879 from the Space Telescope Science Institute (STScI), which is operated by AURA, Inc., under NASA contract NAS 5-26555.

A.G.R., S.C. and L.M.M. gratefully acknowledge support by the Munich Institute for Astro- and Particle Physics (MIAPP) of the DFG cluster of excellence ``Origin and Structure of the Universe.''   JCZ is supported by an NSF Astronomy and Astrophysics Postdoctoral Fellowship under award AST-2001869.

The {\HST} data used in this paper are available at the MAST archive.
   
\appendix

Listed below are the sources of photometry used to derived transformations from individual observation epochs to mean phase.
  
\begin{deluxetable}{lccccc}
\setlength\tabcolsep{0.4cm}
\def\arraystretch{0.9}
\tablecaption{Ground Data Sources\label{tbl_obj_src}}
\tablenum{3}
\tablehead{
\colhead{Identifier} & \multicolumn5c{References$^a$} \\
\cline{2-6}
& \colhead{Phase determination} & \colhead{$V$} & \colhead{$I$} & \colhead{$J$} & \colhead{$H$}
}
\startdata
AG CRU & 1,22,28,29 & 22,28,29 & 22,28,29 & NA & NA \\
AP PUP & 1,28 & 28 & 28 & NA & NA \\
AP SGR & 1,3-5,7-13,22,28 & 3-5,7-13,22,28 & 11,12,22,28 & NA & NA \\
BF OPH & 1,3-7,17,22,28 & 3-7,22,28 & 22,28 & NA & NA \\
BG VEL & 1,28 & 28 & 28 & NA & NA \\
ER CAR & 25,28 & 28 & 28 & NA & NA \\
R CRU & 1,28 & 28 & 28 & NA & NA \\
R MUS & 25,28 & 28 & 28 & NA & NA \\
R TRA & 1,22,28 & 22,28 & 22,28 & NA & NA \\
RV SCO & 1,3,22,28 & 3,22,28 & 22,28 & NA & NA \\
RX CAM & 3,21,24,28 & 3,21,24,28 & 3 & NA & NA \\
RY CMA & 1,3,28 & 3,28 & 28 & NA & NA \\
S CRU & 1,22,28 & 22,28 & 22,28 & NA & NA \\
S TRA & 1,22,28 & 22,28 & 22,28 & NA & NA \\
SS SCT & 1-4,9,20,22,28 & 3,4,9,20,22 & 22 & NA & NA \\
T CRU & 1,28 & 28 & 28 & NA & NA \\
T VEL & 1,17,28 & 28 & 28 & NA & NA \\
TT AQL & 1-3,8-10,12-16,18,19,28 & 3,8-10,12-16,18,19,28 & 12,19,28 & 2,18 & 2,18 \\
TX CYG & 3,8,9,15,16,24,28 & 3,8,9,15,16,24,28 & 3 & NA & NA \\
U AQL & 1-14,28 & 3-14,28 & 30 & NA & NA \\
U SGR & 1-5,8-13,16,17,22-24,28 & 3-5,8-13,16,22-24,28 & 11,12,22,28 & NA & NA \\
V CAR & 1,17,28 & 28 & 28 & NA & NA \\
V VEL & 1,26,28 & 26,28 & 26,28 & NA & NA \\
V386 CYG & 3,21,23,28 & 3,21,23,28 & 3 & NA & NA \\
V482 SCO & 1,3,22,28 & 3,22,28 & 22,28 & NA & NA \\
V636 SCO & 25,28 & 28 & 28 & NA & NA \\
W GEM & 1,3,14,21,24,28 & 3,14,21,24,27,28 & 30 & NA & NA \\
\enddata
\tablecomments{
$^a$ The labels are described in Table~\ref{tbl_ref_src}. NA indicates no ground data avaliable.\\
}
\setlength\tabcolsep{6pt}
\def\arraystretch{1}
\end{deluxetable}

\begin{deluxetable}{lcc}
\setlength\tabcolsep{0.4cm}
\def\arraystretch{0.9}
\tablecaption{References for the Labels in Table~\ref{tbl_obj_src}\label{tbl_ref_src}}
\tablenum{4}
\tablehead{
\colhead{Reference ID} & \colhead{Reference} & \colhead{Comments}
}
\startdata
1 & \citet{Pel1976} & McMaster\\
2 & \citet{Welch+1984} & McMaster\\
3 & \citet{Moffett+1984} & McMaster\\
4 & \citet{1992AandAT....2..107B} & McMaster\\
5 & \citet{1992AandAT....2..157B} & McMaster\\
6 & \citet{1992AandAT....2....1B} & McMaster\\
7 & \citet{1992AandAT....2...31B} & McMaster\\
8 & \citet{1992AandAT....2...43B} & McMaster\\
9 & \citet{Berdnikov1992} & McMaster\\
10 & \citet{1993AstL...19...84B} & McMaster\\
11 & \citet{Berdnikov+1995} & McMaster\\
12 & \citet{Berdnikov+1995} & McMaster\\
13 & \citet{1995AstL...21..308B} & McMaster\\
14 & \citet{Kiss1998} & McMaster\\
15 & \citet{Szabados1981} & McMaster\\
16 & \citet{1986PZ.....22..369B} & McMaster\\
17 & \citet{Laney+1992} & McMaster\\
18 & \citet{Barnes+1997} & McMaster\\
19 & \citet{Coulson+1985b} & McMaster\\
20 & \citet{Henden1980} & McMaster\\
21 & \citet{Szabados1980} & McMaster\\
22 & \citet{Gieren1981} & McMaster\\
23 & \citet{Berdnikov1987} & McMaster\\
24 & \citet{Harris1980} & McMaster\\
25 & \citet{Walraven+1964} & McMaster\\
26 & \citet{Gieren1985} & McMaster\\
27 & \citet{2017PASP..129j4502K} & ASAS-SN\\
28 & \citet{Berdnikov+2000} & \\
29 & \citet{2015yCat..90410027B} & \\
30 & AAVSO & AAVSO\\
\enddata
\setlength\tabcolsep{6pt}
\def\arraystretch{1}
\end{deluxetable}

\clearpage
\bibliographystyle{aasjournal} %
\bibliography{bibdesk}

\begin{thebibliography}{}
\expandafter\ifx\csname natexlab\endcsname\relax\def\natexlab#1{#1}\fi
\providecommand{\url}[1]{\href{#1}{#1}}

\bibitem[{{Anderson} \& {Riess}(2018)}]{Anderson:2018}
{Anderson}, R.~I., \& {Riess}, A.~G. 2018, \apj, 861, 36

\bibitem[{{Arenou} {et~al.}(2018){Arenou}, {Luri}, {Babusiaux}, {Fabricius},
  {Helmi}, {Muraveva}, {Robin}, {Spoto}, {Vallenari}, {Antoja},
  {Cantat-Gaudin}, {Jordi}, {Leclerc}, {Reyl{\'e}}, {Romero-G{\'o}mez}, {Shih},
  {Soria}, {Barache}, {Bossini}, {Bragaglia}, {Breddels}, {Fabrizio},
  {Lambert}, {Marrese}, {Massari}, {Moitinho}, {Robichon}, {Ruiz-Dern},
  {Sordo}, {Veljanoski}, {Di Matteo}, {Eyer}, {Jasniewicz}, {Pancino},
  {Soubiran}, {Spagna}, {Tanga}, {Turon}, \& {Zurbach}}]{Arenou:2018}
{Arenou}, F., {Luri}, X., {Babusiaux}, C., {et~al.} 2018, ArXiv e-prints,
  arXiv:1804.09375

\bibitem[{{Barnes} {et~al.}(1997){Barnes}, {Fernley}, {Frueh}, {Navas},
  {Moffett}, \& {Skillen}}]{Barnes+1997}
{Barnes}, III, T.~G., {Fernley}, J.~A., {Frueh}, M.~L., {et~al.} 1997, \pasp,
  109, 645

\bibitem[{{Berdnikov}(1986)}]{1986PZ.....22..369B}
{Berdnikov}, L.~N. 1986, Peremennye Zvezdy, 22, 369

\bibitem[{{Berdnikov}(1987)}]{Berdnikov1987}
---. 1987, Peremennye Zvezdy, 22, 530

\bibitem[{{Berdnikov}(1992{\natexlab{a}})}]{1992AandAT....2..107B}
---. 1992{\natexlab{a}}, Astronomical and Astrophysical Transactions, 2, 107

\bibitem[{{Berdnikov}(1992{\natexlab{b}})}]{1992AandAT....2..157B}
---. 1992{\natexlab{b}}, Astronomical and Astrophysical Transactions, 2, 157

\bibitem[{{Berdnikov}(1992{\natexlab{c}})}]{1992AandAT....2....1B}
---. 1992{\natexlab{c}}, Astronomical and Astrophysical Transactions, 2, 1

\bibitem[{{Berdnikov}(1992{\natexlab{d}})}]{1992AandAT....2...31B}
---. 1992{\natexlab{d}}, Astronomical and Astrophysical Transactions, 2, 31

\bibitem[{{Berdnikov}(1992{\natexlab{e}})}]{1992AandAT....2...43B}
---. 1992{\natexlab{e}}, Astronomical and Astrophysical Transactions, 2, 43

\bibitem[{{Berdnikov}(1992{\natexlab{f}})}]{Berdnikov1992}
---. 1992{\natexlab{f}}, Soviet Astronomy Letters, 18, 130

\bibitem[{{Berdnikov}(1993)}]{1993AstL...19...84B}
---. 1993, Astronomy Letters, 19, 84

\bibitem[{{Berdnikov} {et~al.}(2000){Berdnikov}, {Dambis}, \&
  {Vozyakova}}]{Berdnikov+2000}
{Berdnikov}, L.~N., {Dambis}, A.~K., \& {Vozyakova}, O.~V. 2000, \aaps, 143,
  211

\bibitem[{{Berdnikov} {et~al.}(2015){Berdnikov}, {Kniazev}, {Sefako}, {Dambis},
  {Kravtsov}, \& {Zhuiko}}]{2015yCat..90410027B}
{Berdnikov}, L.~N., {Kniazev}, A.~Y., {Sefako}, R., {et~al.} 2015, VizieR
  Online Data Catalog, 904

\bibitem[{{Berdnikov} \& {Turner}(1995)}]{Berdnikov+1995}
{Berdnikov}, L.~N., \& {Turner}, D.~G. 1995, Astronomy Letters, 21, 717

\bibitem[{{Berdnikov} \& {Vozyakova}(1995)}]{1995AstL...21..308B}
{Berdnikov}, L.~N., \& {Vozyakova}, O.~V. 1995, Astronomy Letters, 21, 308

\bibitem[{{Breuval} {et~al.}(2020){Breuval}, {Kervella}, {Anderson}, {Riess},
  {Arenou}, {Trahin}, {M{\'e}rand}, {Gallenne}, {Gieren}, {Storm}, {Bono},
  {Pietrzy{\'n}ski}, {Nardetto}, {Javanmardi}, \& {Hocd{\'e}}}]{Breuval:2020}
{Breuval}, L., {Kervella}, P., {Anderson}, R.~I., {et~al.} 2020, arXiv
  e-prints, arXiv:2006.08763

\bibitem[{{Butkevich} {et~al.}(2017){Butkevich}, {Klioner}, {Lindegren},
  {Hobbs}, \& {van Leeuwen}}]{Butkevich:2017}
{Butkevich}, A.~G., {Klioner}, S.~A., {Lindegren}, L., {Hobbs}, D., \& {van
  Leeuwen}, F. 2017, \aap, 603, A45

\bibitem[{{Coulson} {et~al.}(1985){Coulson}, {Caldwell}, \&
  {Gieren}}]{Coulson+1985b}
{Coulson}, I.~M., {Caldwell}, J.~A.~R., \& {Gieren}, W.~P. 1985, \apjs, 57, 595

\bibitem[{{Evans}(1995)}]{Evans:1995}
{Evans}, N.~R. 1995, \apj, 445, 393

\bibitem[{{Fabricius} {et~al.}(2020){Fabricius}, {Luri}, {Arenou}, {Babusiaux},
  {Helmi}, {Muraveva}, {Reyl{\'e}}, {Spoto}, {Vallenari}, {Antoja}, {Balbinot},
  {Barache}, {Bauchet}, {Bragaglia}, {Busonero}, {Cantat-Gaudin}, {Carrasco},
  {Diakit{\'e}}, {Fabrizio}, {Figueras}, {Garcia-Gutierrez}, {Garofalo},
  {Jordi}, {Kervella}, {Khanna}, {Leclerc}, {Licata}, {Lambert}, {Marrese},
  {Masip}, {Ramos}, {Robichon}, {Robin}, {Romero-G{\'o}mez}, {Rubele}, \&
  {Weiler}}]{Fabricius:2020}
{Fabricius}, C., {Luri}, X., {Arenou}, F., {et~al.} 2020, arXiv e-prints,
  arXiv:2012.06242

\bibitem[{{Gaia Collaboration} {et~al.}(2018){Gaia Collaboration}, {Brown},
  {Vallenari}, {Prusti}, {de Bruijne}, {Babusiaux}, \&
  {Bailer-Jones}}]{Gaia-Collaboration:2018}
{Gaia Collaboration}, {Brown}, A.~G.~A., {Vallenari}, A., {et~al.} 2018, ArXiv
  e-prints, arXiv:1804.09365

\bibitem[{{Gaia Collaboration} {et~al.}(2020){Gaia Collaboration}, {Brown},
  {Vallenari}, {Prusti}, {de Bruijne}, {Babusiaux}, \&
  {Biermann}}]{Gaia-Collaboration:2020}
---. 2020, arXiv e-prints, arXiv:2012.01533

\bibitem[{{Gaia Collaboration} {et~al.}(2016{\natexlab{a}}){Gaia
  Collaboration}, {Prusti}, {de Bruijne}, {Brown}, {Vallenari}, {Babusiaux},
  {Bailer-Jones}, {Bastian}, {Biermann}, {Evans}, \&
  et~al.}]{Gaia-Collaboration:2016}
{Gaia Collaboration}, {Prusti}, T., {de Bruijne}, J.~H.~J., {et~al.}
  2016{\natexlab{a}}, \aap, 595, A1

\bibitem[{{Gaia Collaboration} {et~al.}(2016{\natexlab{b}}){Gaia
  Collaboration}, {Brown}, {Vallenari}, {Prusti}, {de Bruijne}, {Mignard},
  {Drimmel}, {Babusiaux}, {Bailer-Jones}, {Bastian}, \&
  et~al.}]{Gaia-Collaboration:2016a}
{Gaia Collaboration}, {Brown}, A.~G.~A., {Vallenari}, A., {et~al.}
  2016{\natexlab{b}}, \aap, 595, A2

\bibitem[{{Gallenne} {et~al.}(2019){Gallenne}, {Kervella}, {Borgniet},
  {M{\'e}rand}, {Pietrzy{\'n}ski}, {Gieren}, {Monnier}, {Schaefer}, {Evans},
  {Anderson}, {Baron}, {Roettenbacher}, \& {Karczmarek}}]{Gallenne:2019}
{Gallenne}, A., {Kervella}, P., {Borgniet}, S., {et~al.} 2019, \aap, 622, A164

\bibitem[{{Gieren}(1981)}]{Gieren1981}
{Gieren}, W. 1981, \apjs, 47, 315

\bibitem[{{Gieren}(1985)}]{Gieren1985}
{Gieren}, W.~P. 1985, \apj, 295, 507

\bibitem[{{Groenewegen}(2018)}]{Groenwegen:2018}
{Groenewegen}, M.~A.~T. 2018, \aap, 619, A8

\bibitem[{{Harris}(1980)}]{Harris1980}
{Harris}, H.~C. 1980, PhD thesis, Washington Univ., Seattle.

\bibitem[{{Henden}(1980)}]{Henden1980}
{Henden}, A.~A. 1980, \mnras, 192, 621

\bibitem[{{Hoffmann} {et~al.}(2016){Hoffmann}, {Macri}, {Riess}, {Yuan},
  {Casertano}, {Foley}, {Filippenko}, {Tucker}, {Chornock}, {Silverman},
  {Welch}, {Goobar}, \& {Amanullah}}]{Hoffmann:2016}
{Hoffmann}, S.~L., {Macri}, L.~M., {Riess}, A.~G., {et~al.} 2016, \apj, 830, 10

\bibitem[{{Kervella} {et~al.}(2019){Kervella}, {Gallenne}, {Evans}, {Szabados},
  {Arenou}, {M{\'e}rand}, {Nardetto}, {Gieren}, \&
  {Pietrzynski}}]{Kervella:2019}
{Kervella}, P., {Gallenne}, A., {Evans}, N.~R., {et~al.} 2019, \aap, 623, A117

\bibitem[{{Kiss}(1998)}]{Kiss1998}
{Kiss}, L.~L. 1998, \mnras, 297, 825

\bibitem[{{Kochanek} {et~al.}(2017){Kochanek}, {Shappee}, {Stanek}, {Holoien},
  {Thompson}, {Prieto}, {Dong}, {Shields}, {Will}, {Britt}, {Perzanowski}, \&
  {Pojma{\'n}ski}}]{2017PASP..129j4502K}
{Kochanek}, C.~S., {Shappee}, B.~J., {Stanek}, K.~Z., {et~al.} 2017, \pasp,
  129, 104502

\bibitem[{{Laney} \& {Stobie}(1992)}]{Laney+1992}
{Laney}, C.~D., \& {Stobie}, R.~S. 1992, \aaps, 93, 93

\bibitem[{{Lindegren} {et~al.}(2018){Lindegren}, {Hernandez}, {Bombrun},
  {Klioner}, {Bastian}, {Ramos-Lerate}, {de Torres}, {Steidelmuller},
  {Stephenson}, {Hobbs}, {Lammers}, {Biermann}, {Geyer}, {Hilger}, {Michalik},
  {Stampa}, {McMillan}, {Castaneda}, {Clotet}, {Comoretto}, {Davidson},
  {Fabricius}, {Gracia}, {Hambly}, {Hutton}, {Mora}, {Portell}, {van Leeuwen},
  {Abbas}, {Abreu}, {Altmann}, {Andrei}, {Anglada}, {Balaguer-Nunez},
  {Barache}, {Becciani}, {Bertone}, {Bianchi}, {Bouquillon}, {Bourda},
  {Brusemeister}, {Bucciarelli}, {Busonero}, {Buzzi}, {Cancelliere},
  {Carlucci}, {Charlot}, {Cheek}, {Crosta}, {Crowley}, {de Bruijne}, {de
  Felice}, {Drimmel}, {Esquej}, {Fienga}, {Fraile}, {Gai}, {Garralda},
  {Gonzalez-Vidal}, {Guerra}, {Hauser}, {Hofmann}, {Holl}, {Jordan},
  {Lattanzi}, {Lenhardt}, {Liao}, {Licata}, {Lister}, {Loffler}, {Marchant},
  {Martin-Fleitas}, {Messineo}, {Mignard}, {Morbidelli}, {Poggio}, {Riva},
  {Rowell}, {Salguero}, {Sarasso}, {Sciacca}, {Siddiqui}, {Smart}, {Spagna},
  {Steele}, {Taris}, {Torra}, {van Elteren}, {van Reeven}, \&
  {Vecchiato}}]{Lindegren:2018}
{Lindegren}, L., {Hernandez}, J., {Bombrun}, A., {et~al.} 2018, ArXiv e-prints,
  arXiv:1804.09366

\bibitem[{{Lindegren} {et~al.}(2020{\natexlab{a}}){Lindegren}, {Klioner},
  {Hern{\'a}ndez}, {Bombrun}, {Ramos-Lerate}, {Steidelm{\"u}ller}, {Bastian},
  {Biermann}, {de Torres}, {Gerlach}, {Geyer}, {Hilger}, {Hobbs}, {Lammers},
  {McMillan}, {Stephenson}, {Casta{\~n}eda}, {Davidson}, {Fabricius},
  {Gracia-Abril}, {Portell}, {Rowell}, {Teyssier}, {Torra}, {Bartolom{\'e}},
  {Clotet}, {Garralda}, {Gonz{\'a}lez-Vidal}, {Torra}, {Abbas}, {Altmann},
  {Anglada Varela}, {Balaguer-N{\'u}{\~n}ez}, {Balog}, {Barache}, {Becciani},
  {Bernet}, {Bertone}, {Bianchi}, {Bouquillon}, {Brown}, {Bucciarelli},
  {Busonero}, {Butkevich}, {Buzzi}, {Cancelliere}, {Carlucci}, {Charlot},
  {Cioni}, {Crosta}, {Crowley}, {del Peloso}, {del Pozo}, {Drimmel}, {Esquej},
  {Fienga}, {Fraile}, {Gai}, {Garcia-Reinaldos}, {Guerra}, {Hambly}, {Hauser},
  {Jan{\ss}en}, {Jordan}, {Kostrzewa-Rutkowska}, {Lattanzi}, {Liao}, {Licata},
  {Lister}, {L{\"o}ffler}, {Marchant}, {Masip}, {Mignard}, {Mints}, {Molina},
  {Mora}, {Morbidelli}, {Murphy}, {Pagani}, {Panuzzo}, {Pe{\~n}alosa Esteller},
  {Poggio}, {Re Fiorentin}, {Riva}, {Sagrist{\`a} Sell{\'e}s}, {Sanchez
  Gimenez}, {Sarasso}, {Sciacca}, {Siddiqui}, {Smart}, {Souami}, {Spagna},
  {Steele}, {Taris}, {Utrilla}, {van Reeven}, \& {Vecchiato}}]{Lindegren:2020a}
{Lindegren}, L., {Klioner}, S.~A., {Hern{\'a}ndez}, J., {et~al.}
  2020{\natexlab{a}}, arXiv e-prints, arXiv:2012.03380

\bibitem[{{Lindegren} {et~al.}(2020{\natexlab{b}}){Lindegren}, {Bastian},
  {Biermann}, {Bombrun}, {de Torres}, {Gerlach}, {Geyer}, {Hern{\'a}ndez},
  {Hilger}, {Hobbs}, {Klioner}, {Lammers}, {McMillan}, {Ramos-Lerate},
  {Steidelm{\"u}ller}, {Stephenson}, \& {van Leeuwen}}]{Lindegren:2020b}
{Lindegren}, L., {Bastian}, U., {Biermann}, M., {et~al.} 2020{\natexlab{b}},
  arXiv e-prints, arXiv:2012.01742

\bibitem[{{Macri} {et~al.}(2015){Macri}, {Ngeow}, {Kanbur}, {Mahzooni}, \&
  {Smitka}}]{Macri:2015}
{Macri}, L.~M., {Ngeow}, C.-C., {Kanbur}, S.~M., {Mahzooni}, S., \& {Smitka},
  M.~T. 2015, \aj, 149, 117

\bibitem[{{Madore}(1982)}]{madore82}
{Madore}, B.~F. 1982, \apj, 253, 575

\bibitem[{{Moffett} \& {Barnes}(1984)}]{Moffett+1984}
{Moffett}, T.~J., \& {Barnes}, III, T.~G. 1984, \apjs, 55, 389

\bibitem[{{Pel}(1976)}]{Pel1976}
{Pel}, J.~W. 1976, \aaps, 24, 413

\bibitem[{{Persson} {et~al.}(2004){Persson}, {Madore}, {Krzemi{\'n}ski},
  {Freedman}, {Roth}, \& {Murphy}}]{persson04}
{Persson}, S.~E., {Madore}, B.~F., {Krzemi{\'n}ski}, W., {et~al.} 2004, \aj,
  128, 2239

\bibitem[{{Riess} {et~al.}(2019){Riess}, {Narayan}, \& {Calamida}}]{Riess:2019}
{Riess}, A.~G., {Narayan}, G., \& {Calamida}, A. 2019, {Calibration of the
  WFC3-IR Count-rate Nonlinearity, Sub-percent Accuracy for a Factor of a
  Million in Flux}, Tech. rep.

\bibitem[{{Riess} {et~al.}(2016){Riess}, {Macri}, {Hoffmann}, {Scolnic},
  {Casertano}, {Filippenko}, {Tucker}, {Reid}, {Jones}, {Silverman},
  {Chornock}, {Challis}, {Yuan}, {Brown}, \& {Foley}}]{Riess:2016}
{Riess}, A.~G., {Macri}, L.~M., {Hoffmann}, S.~L., {et~al.} 2016, \apj, 826, 56

\bibitem[{{Riess} {et~al.}(2018{\natexlab{a}}){Riess}, {Casertano}, {Yuan},
  {Macri}, {Anderson}, {MacKenty}, {Bowers}, {Clubb}, {Filippenko}, {Jones}, \&
  {Tucker}}]{Riess:2018a}
{Riess}, A.~G., {Casertano}, S., {Yuan}, W., {et~al.} 2018{\natexlab{a}}, \apj,
  855, 136

\bibitem[{{Riess} {et~al.}(2018{\natexlab{b}}){Riess}, {Casertano}, {Yuan},
  {Macri}, {Bucciarelli}, {Lattanzi}, {MacKenty}, {Bowers}, {Zheng},
  {Filippenko}, {Huang}, \& {Anderson}}]{Riess:2018b}
---. 2018{\natexlab{b}}, \apj, 861, 126

\bibitem[{{Sahu} {et~al.}(2015){Sahu}, {Gosmeyer}, \& {Baggett}}]{Sahu:2015}
{Sahu}, K., {Gosmeyer}, C.~M., \& {Baggett}, S. 2015, {WFC3/UVIS Shutter
  Characterization}, Tech. rep.

\bibitem[{{Scolnic} {et~al.}(2018){Scolnic}, {Jones}, {Rest}, {Pan},
  {Chornock}, {Foley}, {Huber}, {Kessler}, {Narayan}, {Riess}, {Rodney},
  {Berger}, {Brout}, {Challis}, {Drout}, {Finkbeiner}, {Lunnan}, {Kirshner},
  {Sanders}, {Schlafly}, {Smartt}, {Stubbs}, {Tonry}, {Wood-Vasey}, {Foley},
  {Hand}, {Johnson}, {Burgett}, {Chambers}, {Draper}, {Hodapp}, {Kaiser},
  {Kudritzki}, {Magnier}, {Metcalfe}, {Bresolin}, {Gall}, {Kotak}, {McCrum}, \&
  {Smith}}]{Scolnic:2018}
{Scolnic}, D.~M., {Jones}, D.~O., {Rest}, A., {et~al.} 2018, \apj, 859, 101

\bibitem[{{Szabados}(1980)}]{Szabados1980}
{Szabados}, L. 1980, Commmunications of the Konkoly Observatory Hungary, 76, 1

\bibitem[{{Szabados}(1981)}]{Szabados1981}
---. 1981, Commmunications of the Konkoly Observatory Hungary, 77, 1

\bibitem[{{Szabados} {et~al.}(2013){Szabados}, {Derekas}, {Kiss}, {Kov{\'a}cs},
  {Anderson}, {Kiss}, {Szalai}, {Sz{\'e}kely}, \&
  {Christiansen}}]{Szabados:2013}
{Szabados}, L., {Derekas}, A., {Kiss}, L.~L., {et~al.} 2013, \mnras, 430, 2018

\bibitem[{{Verde} {et~al.}(2019){Verde}, {Treu}, \& {Riess}}]{Verde:2019}
{Verde}, L., {Treu}, T., \& {Riess}, A.~G. 2019, Nature Astronomy, 3, 891

\bibitem[{{Walraven} {et~al.}(1964){Walraven}, {Tinbergen}, \&
  {Walraven}}]{Walraven+1964}
{Walraven}, J.~H., {Tinbergen}, J., \& {Walraven}, T. 1964, \bain, 17, 520

\bibitem[{{Welch} {et~al.}(1984){Welch}, {Wieland}, {McAlary}, {McGonegal},
  {Madore}, {McLaren}, \& {Neugebauer}}]{Welch+1984}
{Welch}, D.~L., {Wieland}, F., {McAlary}, C.~W., {et~al.} 1984, \apjs, 54, 547

\bibitem[{{Zinn}(2020)}]{Zinn:2020}
{Zinn}, J.~C. 2020, Characterization of the EDR3 Parallax Offset from Red Giant
  Astero-Seismology, ,

\bibitem[{{Zinn} {et~al.}(2018){Zinn}, {Pinsonneault}, {Huber}, \&
  {Stello}}]{Zinn:2018}
{Zinn}, J.~C., {Pinsonneault}, M.~H., {Huber}, D., \& {Stello}, D. 2018, ArXiv
  e-prints, arXiv:1805.02650

\end{thebibliography}
\clearpage

\begin{figure}[ht]
\vspace*{150mm}
\figurenum{1}
\includegraphics{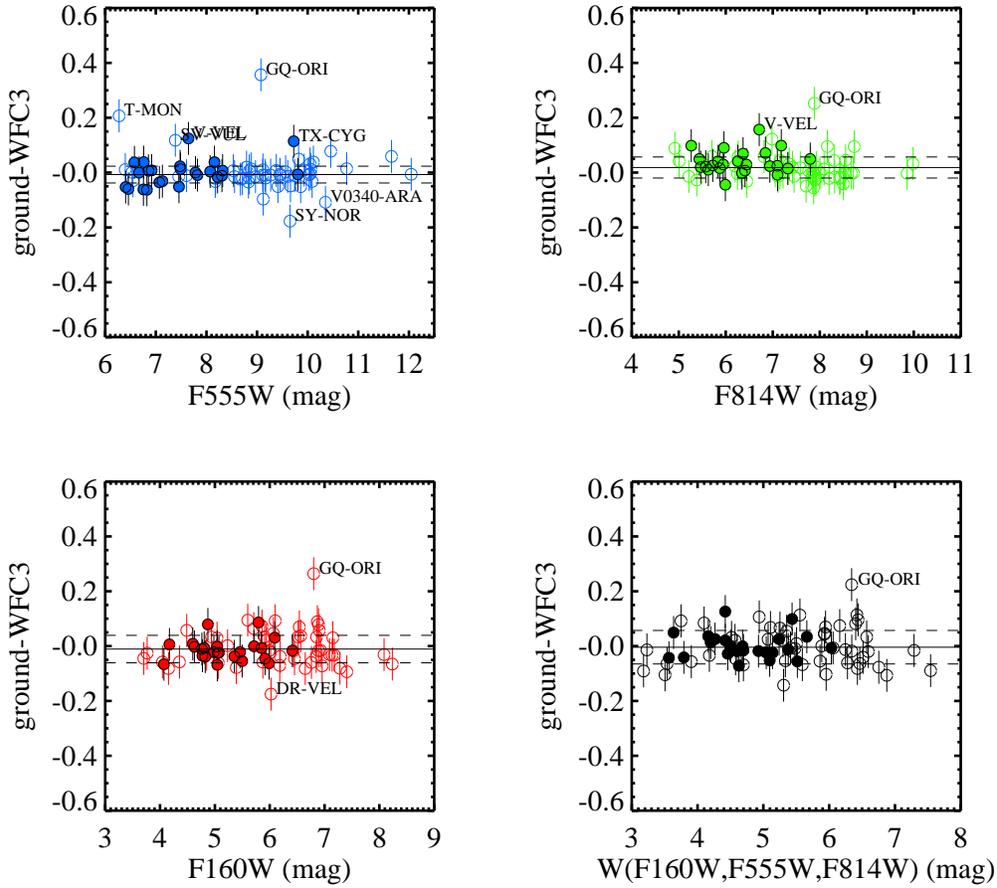}
\caption{\label{fg:outlier}  Comparison of Cepheid mean magnitudes in three {\it HST} WFC3 bands for observations obtained with {\it HST} and from the ground (transformed to the {\it HST} system). Filled circles are from the new Cycle 27 sample, open are from Cycle 22 and R18b.}
\end{figure}

% made with comp_grnd_matts in directory fast_scans

\begin{figure}[ht]
\vspace*{150mm}
\figurenum{2}
\includegraphics{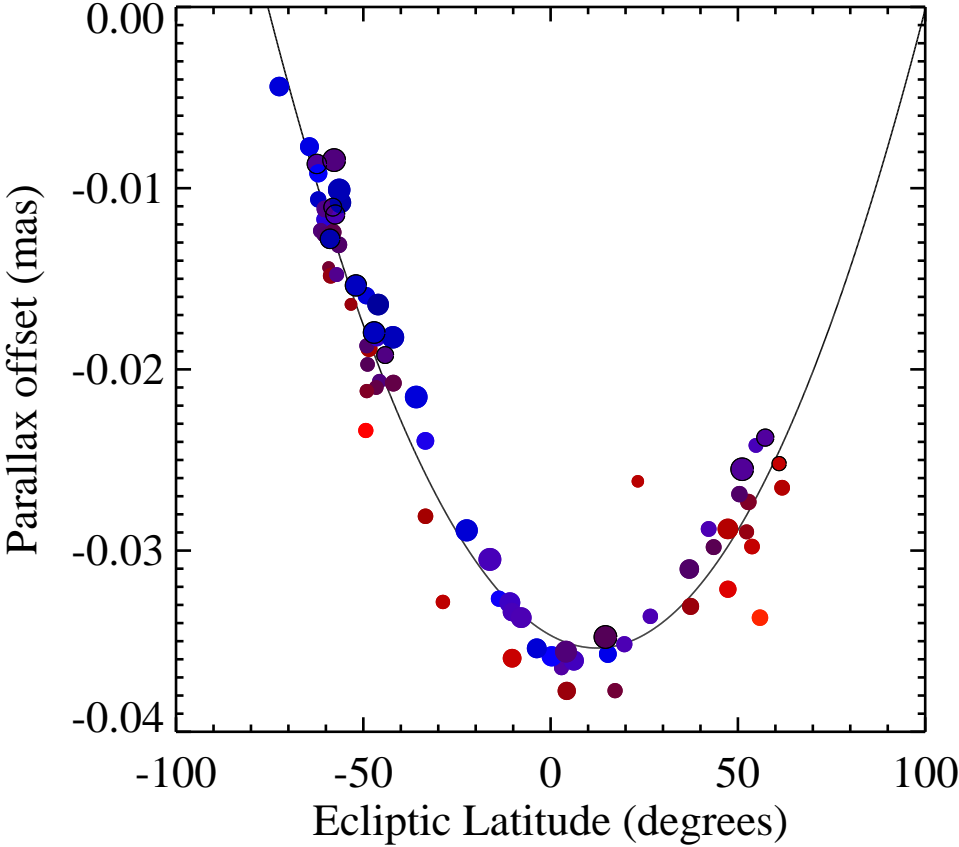}
\caption{\label{fg:outlier}  Parallax offsets from the L20b formulae.  The primary dependence of the parallax offset is a parabolic dependence on ecliptic latitude.  There is a small variation with G magnitude (indicated by the size of the point) and color (indicated by the color of the point).  The Cepheid CY Aur sits about 10 \mias above the parabola and its excluded from the analysis due to an uncertain parallax offset. Circles with black outline have 6-parameter solutions in Gaia EDR3.}
\end{figure}

% made with show_zp in megashoes

\begin{figure*}[ht]
\figurenum{3}
% \vspace*{100mm}
\noindent \vbox{
  \noindent \hbox to \hsize{\hss
    \resizebox{5.5in}{!}{\includegraphics[angle=90]{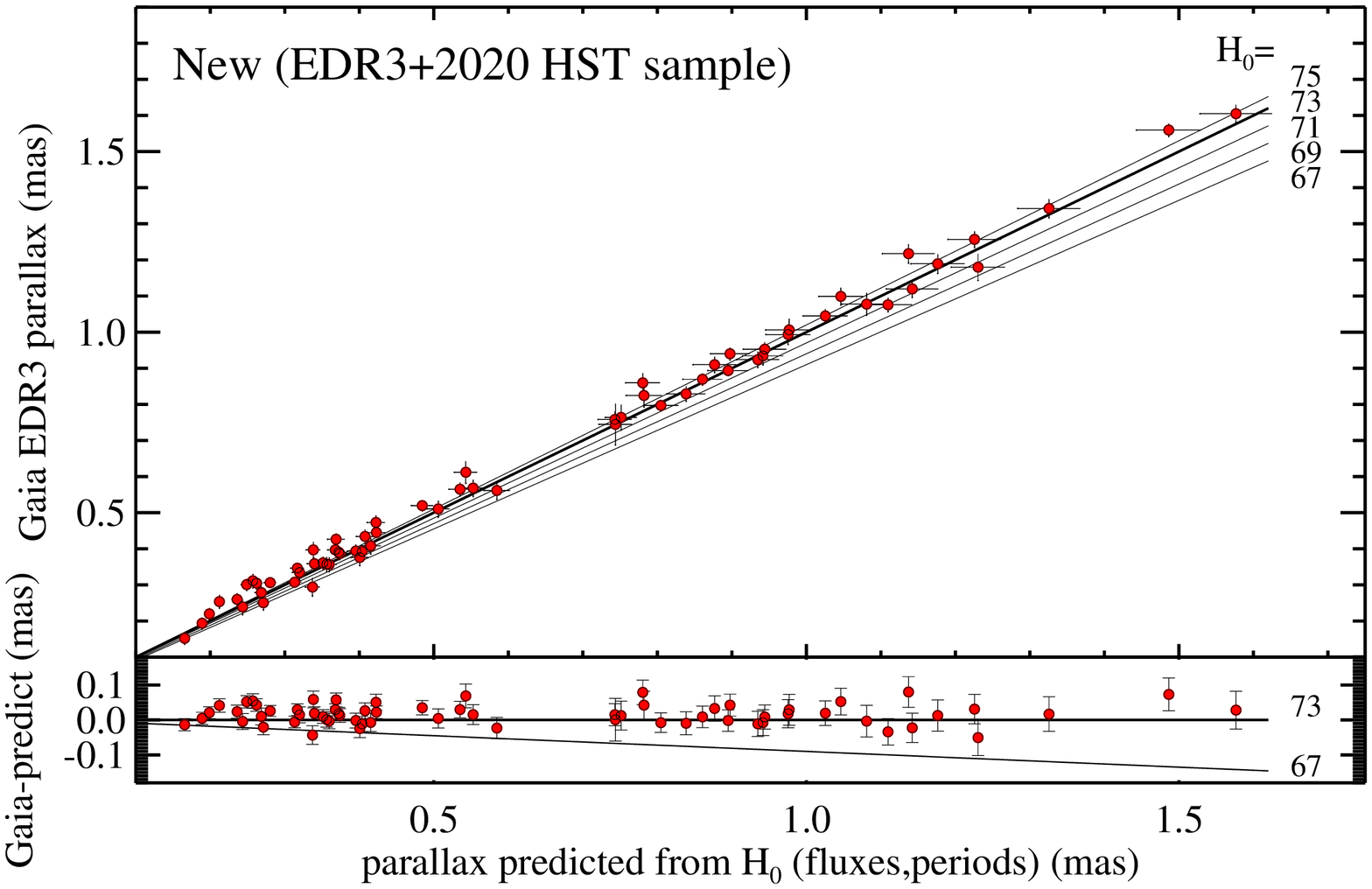}}\hss}
  \vspace {10pt}
  \noindent\hbox to \hsize{\hss
  \resizebox{5.5in}{!}{\includegraphics[angle=90]{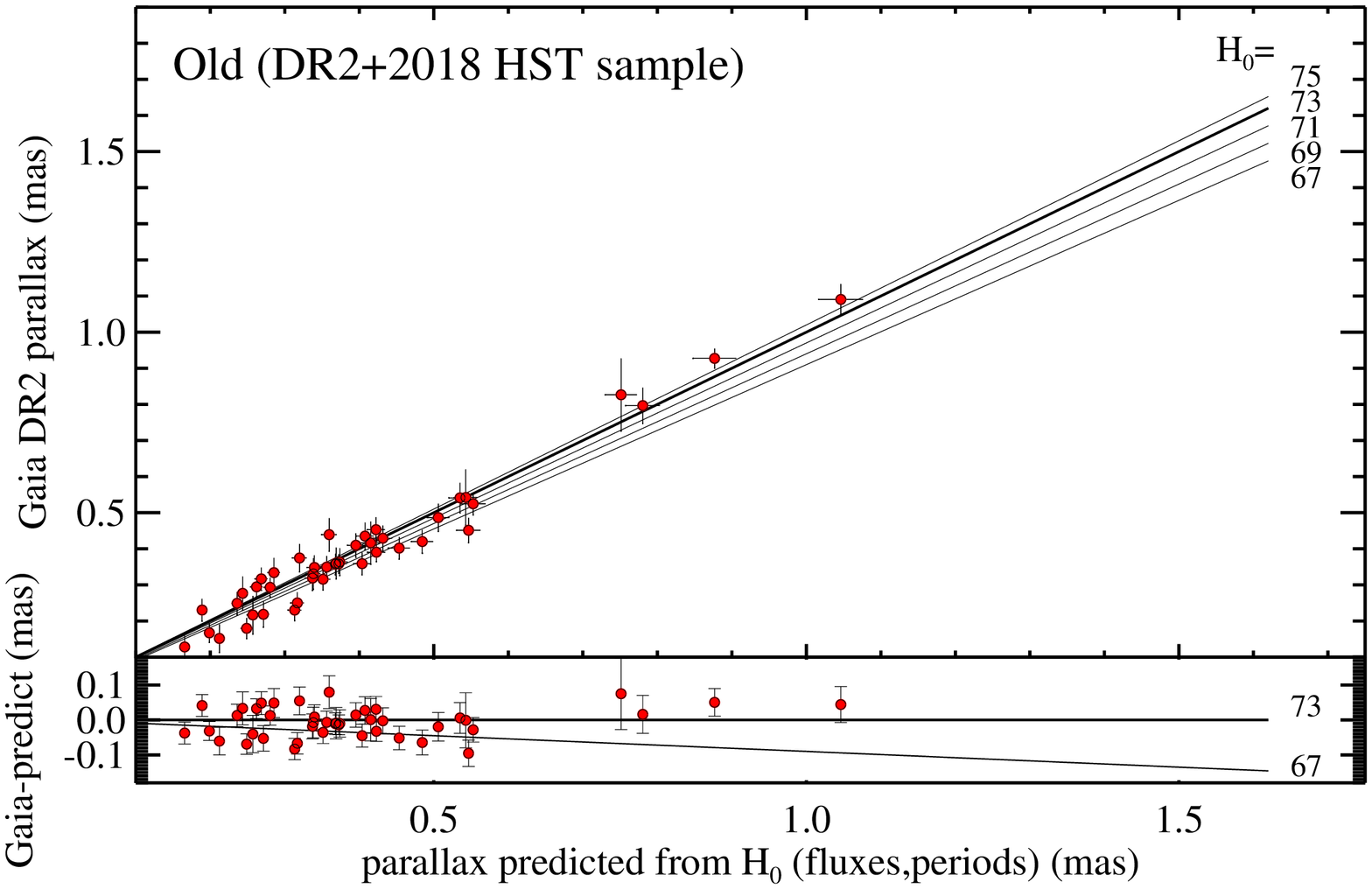}}\hss}
}
\caption{\label{fg:outlier}  Comparison of Milky Way Cepheid parallaxes provided in Gaia EDR3 for the present expanded sample (left)
and DR2 for the earlier, smaller sample available in \citet{Riess:2018b} (right) vs. photometric parallaxes using the {\it HST} WFC3-based photometry in Table 1, 
the Cepheid periods, and the {\PL} parameters given by R16 and R19.  }
\end{figure*}

\begin{figure}[ht]
\vspace*{150mm}
\figurenum{4}
\includegraphics{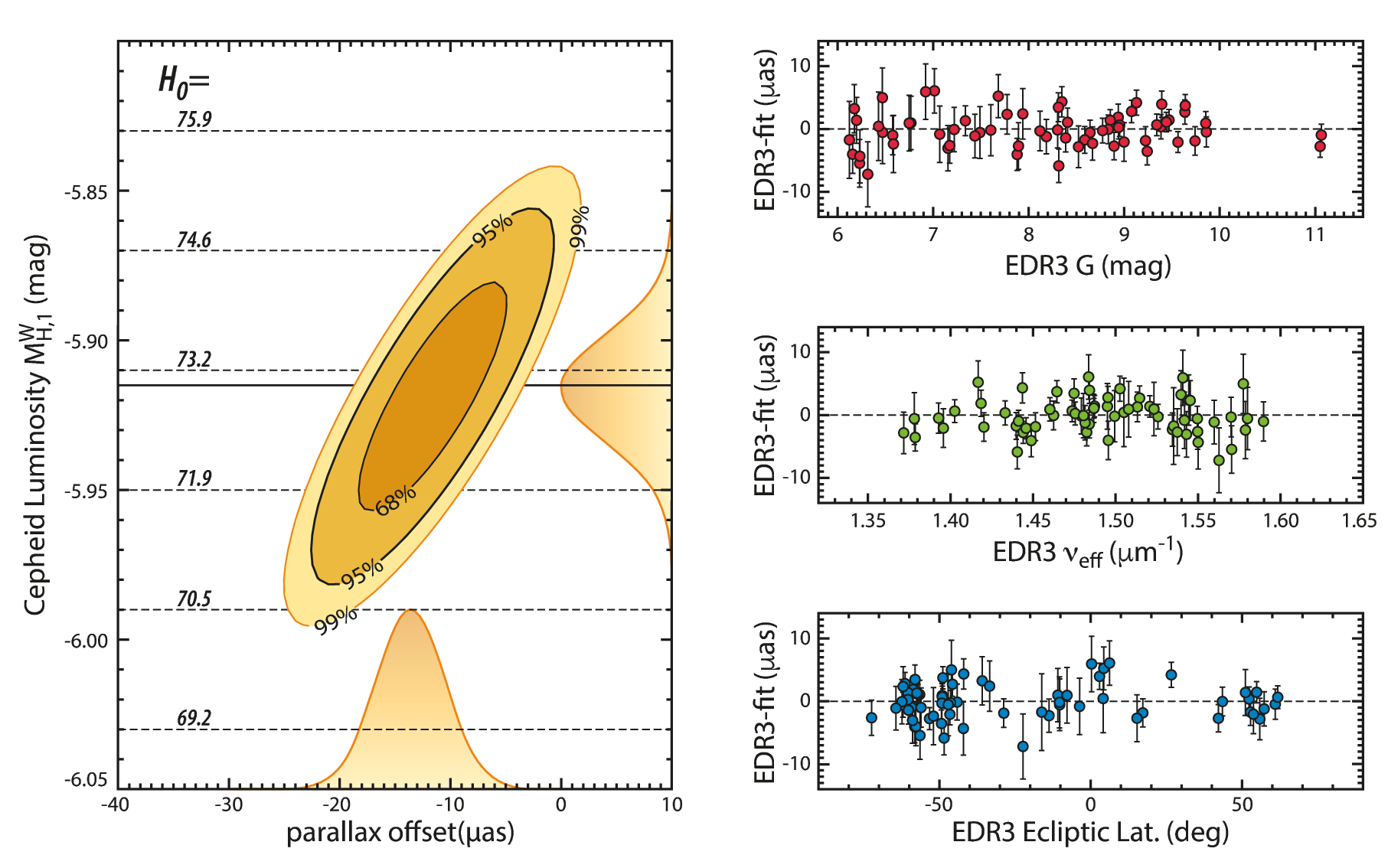}
\caption{\label{fg:outlier}  For the {\it HST} sample of  Milky Way Cepheids, we determined the best match between the measured Gaia EDR3 parallaxes and those predicted {\it photometrically} from their photometry, periods, and the fiducial Cepheid luminosity, $M_{H,1}^W$.  We allow two free parameters, an additive term to parallax to account for the parallax offset, $zp$, and a multiplicative term to predicted parallax that measures the fiducial Cepheid luminosity.  The Cepheid luminosity calibrates  the SH0ES distance ladder from R16 and R19 and results in the indicated values of $H_0$. Right, top to bottom, residuals between the best fit vs. Cepheid G mag, color, and ecliptic latitude.}
\end{figure}

% made with quick_par in megashoes

\end{document}